\documentclass{article}

\usepackage{hyperref}
\usepackage{amssymb}

\usepackage{fullpage}

\usepackage{tikz} 
\usetikzlibrary{backgrounds}
\usetikzlibrary{calc}

\definecolor{grey}{rgb}{0.5,0.5,0.5}
\definecolor{lightgrey}{rgb}{0.7,0.7,0.7}
\definecolor{verylightgrey}{rgb}{0.9,0.9,0.9}
\definecolor{darkgrey}{rgb}{0.2,0.2,0.2}

\definecolor{grey5}{rgb}{0.9,0.9,0.9}
\definecolor{grey4}{rgb}{0.833,0.833,0.833}
\definecolor{grey3}{rgb}{0.766,0.766,0.766}
\definecolor{grey2}{rgb}{0.7,0.7,0.7}
\definecolor{grey1}{rgb}{0.633,0.633,0.633}

\usepackage{natbib}
\bibpunct{(}{)}{;}{a}{}{,}

\title{Relational hyperevent models for polyadic interaction networks}

 \author{J\"urgen Lerner\\
   University of Konstanz, Germany\\
   RWTH Aachen, Germany\\
   \texttt{juergen.lerner@uni-konstanz.de}
         \and
         Alessandro Lomi\\
         University of the Italian Switzerland, Lugano, CH.\\
         \texttt{alessandro.lomi@usi.ch}
 }

 \date{}

\begin{document}
\maketitle

\begin{center}\textit{\footnotesize (preprint of a submitted manuscript)}\end{center}

\begin{abstract}
Polyadic, or ``multicast'' social interaction networks arise when one sender addresses multiple receivers simultaneously. Currently available relational event models (REM) are not well suited to the analysis of polyadic interaction networks because they specify event rates for sets of receivers as functions of dyadic covariates associated with the sender and one receiver at a time. Relational hyperevent models (RHEM) address this problem by specifying event rates as functions of hyperedge covariates associated with the sender and the entire set of receivers. For instance, hyperedge covariates can express the tendency of senders to repeatedly address the same pairs (or larger sets) of receivers -- a simple and frequent pattern in polyadic interaction data which, however, cannot be expressed with dyadic covariates. In this article we demonstrate the potential benefits of RHEMs for the analysis of polyadic social interaction. We define and discuss practically relevant effects that are not available for REMs but may be incorporated in empirical specifications of RHEM. We illustrate the empirical value of RHEM, and compare them with related REM, in a reanalysis of the canonical \textit{Enron} email data.
\end{abstract}

\section{Introduction}
\label{sec:intro}
Data generated by technology-mediated communication typically take the form of sequences of time stamped interaction events involving two or more actors simultaneously. This kind of one-to-many (or ``multicast'') interaction is common. For instance, email users can send messages to any number of receivers \citep{zhou2007strategies,perry2013point}. In this paper we call a social interaction process \textit{polyadic} when it is characterized by events in which one ``sender'' may address multiple ``receivers'' simultaneously. 

Data produced by time-stamped polyadic interaction processes are not unique to social network data produced by technology-mediated interaction (e.\,g., email messaging) or social media (e.\,g., twitter). Polyadic interaction data are encountered frequently in empirical research across the social sciences. Examples include scientific papers citing several references \citep{radicchi2012citation}, courts judgments citing multiple legal precedents \citep{fowler2007network}, patents approved by the patent office citing multiple prior patents \citep{verspagen2007mapping,kuhn2020patent}, and infected persons transmitting a virus to several others simultaneously through group contact  \citep{colizza2007modeling,hancean2021role}. Face-to-face conversations where one speaker addresses multiple alters simultaneously also illustrate the empirical extension of polyadic interaction processes \citep{gibson2005taking}. Polyadic interaction may be directed or undirected. Examples of undirected polyadic interaction networks include meetings attended by multiple participants \citep{freeman2003finding,lerner2021dynamic}, coauthors jointly publishing a paper \citep{newman2004coauthorship}, coordination in task-oriented teams \citep{ahmadpoor2019decoding,guimera2005team,leenders2016once}, groups of countries agreeing to sign a multilateral treaty \citep{hollway2016multilevel,simmons2005constraining}, and class action lawsuits where the plaintiff is a group of people simultaneously bringing a suit to one defendant \citep{bronsteen2002class}.

Despite their obvious heterogeneity, these empirical examples share two defining features. First, interaction among the agents takes the form of time stamped relational events -- rather than relational states such as ``being a friend of,''  ``seeking advice from,'' or ``regularly communicate with.'' Second, the stream of observed events is not generated only by dyadic interaction, but involves multiple network nodes interacting at single points in time. In recent years, the availability of social interaction data sharing these features has increased significantly due to the diffusion of computer-mediated communication and collaborative technologies, the availability of large-scale databases, and the diffusion of automated data collection technologies \citep{lazer2009social}. However, the availability of statistical models capable of analyzing these data by accounting for their constitutive features has remained limited.

Relational event models (REMs) \citep{butts2008relational,brandes2009networks,perry2013point,lerner2013modeling} provide the most promising framework for the analysis of sequences of social interaction events observed in continuous time \citep{bianchi2022fromties,vu2017relational}. Typical REM for networks of time stamped interaction events specify point processes whose intensities are functions of dyadic covariates. For instance, the intensity of events from sender $i$ to receiver $j$ might depend on the age or gender of $i$ and $j$, on their age difference, on the frequency of previous events from $i$ to $j$ or from $j$ to $i$, or on previous events to or from common third actors. Extant research recognizes the problem posed by polyadic interaction and multicast communication. The solution that is typically offered involves specifying intensities for events in which a sender $i$ sends a message to a set of receivers $J=\{j_1,\dots,j_k\}$ \citep{perry2013point}. However, these intensities are still modeled as functions of dyadic covariates $x(i,j),\,j\in J$, considering the sender and only one receiver at a time. In this way, available models for relational events assume that the multiple dyads simultaneously produced by a single polyadic interaction are either independent \citep{perry2013point}, or pertain to fictional ``collective'' receivers \citep{butts2008relational}, such as the whole ``team'' as a receiver of broadcast messages. While empirically useful, and occasionally justified, neither of these solutions is fully satisfactory.

Recently proposed relational hyperevent models (RHEMs) \citep{lerner2019rem,lerner2021dynamic} generalize REMs by specifying the rate of interaction events from sender $i$ to receiver set $J$ as a function of \emph{hyperedge covariates} $x(i,J)$ -- being a function of the sender and the entire set of receivers -- that cannot necessarily be decomposed into dyadic covariates. As a concrete example, the average number of past interactions that pairs of actors $\{j,j'\}\subseteq J$ have jointly received from $i$ can be expressed as a hyperedge covariate $x(i,J)$ -- but not as a sum of dyadic covariates $x(i,j),\,j\in J$. Generalizing edges in graphs that connect exactly two nodes, hyperedges in a hypergraph can connect any number of nodes \citep{berge1989hypergraphs}. A relational hyperevent is a time stamped event indexed by a hyperedge \citep{lerner2021dynamic}.

While the conceptual transition from REM to RHEM may be intuitive, its analytical and empirical implications need more complete and rigorous articulation. In this paper, our goal is to improve our current understanding by specifying appropriate models for polyadic social interaction processes and then evaluating empirical differences with respect to comparable dyadic specifications.

A model specified strictly in terms of dyadic covariates might be unable to capture higher-order dependencies typically present in network data produced by polyadic interaction. In turn, this may yield misleading estimates of network effects, or result in lower model fit. In this paper we take the view that higher-order dependencies -- when present -- should not merely be considered as an inconvenient feature of the data to be, in the best cases, controlled away. Rather, we argue that such higher-order effects provide a unique opportunity for improving our understanding of the structure and dynamics of social interaction processes, and for developing and testing innovative theories of social interaction.

In this paper we intend to demonstrate how differences between REM and RHEM may be directly relevant for empirical studies analyzing polyadic social interaction processes. We define and discuss practically relevant network effects that can be expressed in RHEMs, but not in dyadic specifications of the event rate afforded by currently available REMs. In the empirical part of the paper, we use the canonical Enron email data \citep{zhou2007strategies} to illustrate how to test for polyadic dependencies in empirical data. Moreover, we assess and compare improvements in model fit implied by adding various, dyadic and hyperedge, covariates.

RHEM defined and discussed in this paper directly build on the work by \citet{perry2013point}, as detailed below. A number of other alternative approaches have been proposed to adapt REM to events with several receivers (or several senders). In the original statement of the model, \citet{butts2008relational} proposed the creation of ``virtual'' nodes to represent sets of receivers (or senders). This approach can be appropriate for specifically chosen subsets (such as the whole team as a receiver of broadcast messages), but representing all possible subsets by virtual nodes becomes quickly unfeasible.  \citet{kim2018hyperedge} propose the hyperedge event model for multicast events. Their model specifies dyadic intensities, associated with one sender and one receiver, as a function of dyadic covariates. These intensities then stochastically determine the sender of the next event and subsequently the receiver set of an interaction by that sender. Their framework does not allow for hyperedge covariates as in RHEM. \citet{mulder2021latent} define a latent variable model for multicast interaction. However, in their model the mean suitability score of a receiver for messages initiated by a given sender is still a dyadic function, dependent on the sender and one receiver at a time.

Relational hyperevent models for undirected hyperevents (e.\,g., meeting events) have been proposed in \citet{lerner2021dynamic}. Earlier, RHEM have been mentioned by \citet{lerner2019rem} who also defined RHEM for directed hyperevents -- but did not analyze directed RHEM in any of their empirical examples. Directed RHEM have been applied to modeling contact elicitation networks by infected persons in \citet{hancean2021role}.

We demonstrate how the computation of hyperedge covariates for RHEMs may be performed with the open source software \texttt{eventnet}. The analysis of the Enron email data with RHEM is explained in a step-by-step tutorial linked from \url{https://github.com/juergenlerner/eventnet/wiki/}, from the data preprocessing and the computation of covariates over to model estimation in \texttt{R}. Thus, the analysis reported in the empirical part of this paper is fully reproducible. The software may be adopted in -- and adapted to future empirical studies involving polyadic social interaction processes.

\section{RHEM and hyperedge covariates for multicast social interaction}
\label{sec:rhem}

\subsection{Background on REM based on dyadic covariates}
\label{sec:REM_dyadic}

We start by recalling the point process models for directed interaction networks, following closely the argument and the notation proposed by \citet{perry2013point}. We start our discussion from their model for the more general case of relational events defined over receiver sets of arbitrary size. As observed in their paper, the theoretical results on the consistency of the maximum partial likelihood inference also apply to the case of polyadic (``multicast'') events. It may be worth repeating that the difference between RHEM and the model from \citet{perry2013point} does not lie in the basic modeling framework, but in generalizing dyadic covariates to hyperedge covariates.

Let $\mathcal{I}$ be a finite set of senders and $\mathcal{J}$ be a finite set of receivers, not necessarily disjoint from $\mathcal{I}$. We denote elements from $\mathcal{A}=\mathcal{I}\cup\mathcal{J}$ as \emph{actors}. For a sender $i\in\mathcal{I}$ and a point in time $t>0$, let $\mathcal{J}_t(i)\subseteq\mathcal{J}$ denote the set of actors that could potentially receive an interaction from $i$ at $t$. If $\mathcal{A}=\mathcal{I}=\mathcal{J}$, then it is often the case that $\mathcal{J}_t(i)=\mathcal{I}\setminus\{i\}$, implying that a sender can send interaction events to everyone but herself -- that is, loops are excluded. For a sender $i$ and receiver $j$, let $x_t(i,j)$ be a $p$-dimensional vector of covariates and $\beta_0$ a $p$-dimensional vector of unknown parameters. For a sender $i$ and a positive integer $L$ (giving the receiver set size), the baseline intensity is denoted by $\overline{\lambda}_t(i,L)$. 

\citet{perry2013point} define a model for counting processes on $\mathbb{R}_+\times\mathcal{I}\times\mathcal{P}(\mathcal{J})$ where the intensity on $(i,J)$, with $i\in\mathcal{I}$ and $J\subseteq \mathcal{J}$, is modeled as
\begin{equation}
  \label{eq:lambda_dyadic}
  \lambda_t(i,J)=\overline{\lambda}_t(i,|J|)\exp\left\{
  \beta_0^{\rm T}\sum_{j\in J}x_t(i,j)
  \right\}
  \prod_{j\in J}{\bf 1}\{j\in\mathcal{J}_t(i)\}\enspace.
\end{equation}
Intuitively, $\lambda_t(i,J)\Delta t$ is the expected number of events that $i$ sends to the receiver set $J$ in the time interval $[t,t+\Delta t)$. The intensity $\lambda_t(i,J)$ is assumed to be the baseline intensity $\overline{\lambda}_t(i,|J|)$ multiplied with the relative rate, $\exp\left\{\beta_0^{\rm T}\sum_{j\in J}x_t(i,j)\right\}$. Thus, the parameter vector $\beta_0$ controls which covariates $\{x_t(i,j);\,j\in J\}$ make $J$ a more or less likely receiver set for interaction events sent by $i$. The covariates in $x_t(i,j)$ can depend on actor characteristics, such as the age of $j$ or the age difference of $i$ and $j$, but they can also depend on the history of the process. For instance, the covariates $x_t(i,j)$ can include a count of the number of past events that $i$ has sent to $j$, before $t$.

  Importantly, the model summarized in (\ref{eq:lambda_dyadic}) assumes that the covariate vectors $\{x_t(i,j),\;j\in J\}$ increase or decrease the rate of events from sender $i$ to the receiver set $J$ \emph{independently of each other}. Equivalent to (\ref{eq:lambda_dyadic}), we can model the intensity $\lambda_t(i,J)$ as the baseline intensity $\overline{\lambda}_t(i,|J|)$ multiplied with the relative rate $\exp\left\{\beta_0^{\rm T}x_t(i,J)\right\}$, where the covariates $x_t(i,J)$ of sender $i$ and \emph{receiver set} $J$ are assumed to have the specific form
  \begin{equation}
    \label{eq:covar_dyadic}
    x_t(i,J)=\sum_{j\in J}x_t(i,j)\enspace.
  \end{equation}
  The equation above implies that in the original model the suitability of $j$ as a receiver of interaction sent by $i$ is assumed to be independent of the other receivers $j'\in J\setminus\{j\}$ of the same interaction -- an assumption that is challenged in our paper. Generalizing this model, RHEM allow covariates $x_t(i,J)$ that do not necessarily decompose into a sum of dyadic covariates of the form $\sum_{j\in J}x_t(i,j)$; compare Section~\ref{sec:rhem_model} below.
  
  Let $(t_1,i_1,J_1),\dots,(t_n,i_n,J_n)$ be the observed sequence of polyadic interactions where $(t,i,J)$ indicates that at time $t$ sender $i$ interacts with receiver set $J$. The model from \citet{perry2013point}, given in (\ref{eq:lambda_dyadic}), leads to the log partial likelihood at $t$ evaluated at $\beta\in\mathbb{R}^p$:
  \begin{equation}
    \label{eq:likelihood_dyadic}
    \log L_t(\beta)=\sum_{t_m\leq t}\left(
    \beta^{\rm T}\sum_{j\in J_m}x_{t_m}(i_m,j)
    -
    \log\left[
      \sum_{J\in {\mathcal{J}_{t_m}(i_m) \choose |J_m|} }\exp\left\{
        \beta^{\rm T}\sum_{j\in J}x_{t_m}(i_m,j)
        \right\}
      \right]
    \right)\enspace,
  \end{equation}
  where for a set $X$ and an integer $L$, we write ${X \choose L}=\{X'\subseteq X\,;\;|X'|=L\}$ for the set of all subsets of size $L$. 
  \citet{perry2013point} prove that, under usual assumptions, parameters maximizing (\ref{eq:likelihood_dyadic}), i.\,e., the maximum partial likelihood estimates (MPLE), are a consistent estimator of $\beta_0$.

\subsection{RHEM}
\label{sec:rhem_model}
Building on the model proposed by \citet{perry2013point}, the definition of RHEM for directed polyadic interaction can be obtained by substituting $x_t(i,J)$ for $\sum_{j\in J}x_t(i,j)$ in (\ref{eq:lambda_dyadic}) and (\ref{eq:likelihood_dyadic}). More precisely, given the notation from Section~\ref{sec:REM_dyadic}, RHEM define a model for counting processes on $\mathbb{R}_+\times\mathcal{I}\times\mathcal{P}(\mathcal{J})$ where the intensity on $(i,J)$, with $i\in\mathcal{I}$ and $J\subseteq \mathcal{J}$, is modeled as
\begin{equation}
  \label{eq:lambda_rhem}
  \lambda_t(i,J)=\overline{\lambda}_t(i,|J|)\exp\left\{
  \beta_0^{\rm T}x_t(i,J)
  \right\}
  {\bf 1}\{J\subseteq\mathcal{J}_t(i)\}\enspace.
\end{equation}
The covariates $x_t(i,J)$ do not necessarily decompose into a sum of dyadic covariates and may depend on exogenous actor-level characteristics or they can depend on the history of the process. Examples for the former include the average age of receivers in $J$, the average absolute age difference between the sender $i$ and the receivers in $J$, or the average absolute age difference between pairs of receivers in $J$.  Examples for the latter (i.\,e., history-dependent) covariates include the number of past interactions that $i$ has sent to receiver set $J$:
\[
\#\{\mbox{interaction } (t',i,J) \mbox{ with } t'<t\}\enspace,
\]
the average number of past interactions that actors $j\in J$ have received from $i$:
\[
\frac{\sum_{j\in J}\#\{\mbox{interaction }(t',i,J') \mbox{ with } t'<t\wedge j\in J'\}}{|J|}\enspace,
\]
or the average number of past interactions that pairs of actors $\{j,j'\}\subseteq J$ have jointly received from $i$:
\[
\frac{\sum_{\{j,j'\}\in {J \choose 2}}\#\{\mbox{interaction }(t',i,J') \mbox{ with } t'<t\wedge j,j'\in J'\}}{{|J| \choose 2}}\enspace.
\]
Note that the first and the third of the above examples for history-dependent covariates do not admit a decomposition of the form (\ref{eq:covar_dyadic}). The second of these examples for history dependent covariates admits a decomposition of the form (\ref{eq:covar_dyadic}) \emph{if} we are willing to additionally allow multiplication of covariates with a constant that depends only on the receiver set size $|J|$. 

Let $(t_1,i_1,J_1),\dots,(t_n,i_n,J_n)$ be the observed sequence of polyadic interactions. The model given in (\ref{eq:lambda_rhem}) leads to the following log partial likelihood at $t$ evaluated at $\beta\in\mathbb{R}^p$
  \begin{equation}
    \label{eq:likelihood_rhem}
    \log L_t(\beta)=\sum_{t_m\leq t}\left(
    \beta^{\rm T}x_{t_m}(i_m,J_m)
    -
    \log\left[
      \sum_{J\in {\mathcal{J}_{t_m}(i_m) \choose |J_m|} }\exp\left\{
        \beta^{\rm T}x_{t_m}(i_m,J)
        \right\}
      \right]
    \right)\enspace.
  \end{equation}
  The arguments from \citet{perry2013point} on the consistency of MPLE of their model also apply to parameters maximizing (\ref{eq:likelihood_rhem}). Note that the only difference between (\ref{eq:likelihood_dyadic}) and (\ref{eq:likelihood_rhem}) is that we allow more general covariates.

\subsection{Case control sampling}
\label{sec:sampling}

The log partial likelihood (\ref{eq:likelihood_rhem}) can no longer be approximated by replacing the sum over all $J$ in the risk set ${\mathcal{J}_{t_m}(i_m) \choose |J_m|}$ with a sum over all receivers in $\mathcal{J}_{t_m}(i_m)$, as it was the case for models specified with dyadic covariates in \citet{perry2013point}. Having to sum over all subsets would lead to excessive, or even intractable, computational runtime for all but the smallest $\mathcal{J}_{t_m}(i_m)$ and $J_m$.

Previous work on REM and RHEM for the analysis of large networks \citep{vpr-remslm-15,lerner2020reliability,lerner2021dynamic} suggested to reduce excessive computational runtime by replacing the risk set with a sampled risk set obtained via case control sampling \citep{bgl-mascdcphm-95,keogh2014nested}. The derivation of the partial likelihood based on sampled risk sets (\ref{eq:likelihood_rhem_sampled}), starting from the partial likelihood using the entire risk set (\ref{eq:likelihood_rhem}), can actually be obtained by adapting the general approach from \citet{bgl-mascdcphm-95} to our special case -- and in the remainder of this subsection we will only briefly indicate the relevant steps. We recall that the main contribution of this paper are the hyperedge covariates and not the particular method to estimate model parameters. We point out that when applying RHEM to data with a very small number of nodes, case control sampling is not needed \citep{bianchi2022fromties, lerner2022dynamic} and parameters may be estimated by maximizing (\ref{eq:likelihood_rhem}). In application settings where the number of nodes is larger, we propose to estimate RHEM parameters by maximizing (\ref{eq:likelihood_rhem_sampled}). 

For a given positive integer $k$ (the number of non-events per event), let $\tilde{\mathcal{R}}_{t}(\mathcal{J}_{t}(i),J,k)\subseteq {\mathcal{J}_{t}(i) \choose |J|}$ be a set of subsets of $\mathcal{J}_{t}(i)$ that is sampled uniformly at random from
\begin{equation}\label{eq:control_set}
\left\{\mathcal{R}\subseteq {\mathcal{J}_{t}(i) \choose |J|}\,;\;J\in \mathcal{R}\wedge |\mathcal{R}|=k+1\right\}\enspace.
\end{equation}
In other words, $\tilde{\mathcal{R}}_{t}(\mathcal{J}_{t}(i),J,k)$ contains the observed receiver set $J$ (the ``case'') plus $k$ alternative receiver sets (the ``controls''), sampled without replacement uniformly and independently at random from  $\{J'\subseteq\mathcal{J}_{t}(i)\,;\;|J'|=|J|\wedge J'\neq J\}$.

Adapting the general approach of \citet{bgl-mascdcphm-95} to our special case, we obtain a counting process on $\mathbb{R}_+\times\mathcal{I}\times\mathcal{P}(\mathcal{J})\times\mathcal{P}[\mathcal{P}(\mathcal{J})]$, where the tuple $(t,i,J,\mathcal{R})$ indicates that at time $t$, sender $i$ interacts with receiver set $J$ and $\mathcal{R}$ is the set that has been sampled from (\ref{eq:control_set}). If $\pi_t(\mathcal{R}|i,J)$ denotes the conditional probability that $\mathcal{R}$ is the sampled risk set at $t$, given that an event on $(i,J)$ occurs at $t$, then this counting process has intensity
\[
  \lambda_t(i,J,\mathcal{R})=\overline{\lambda}_t(i,|J|)\exp\left\{
  \beta_0^{\rm T}x_t(i,J)
  \right\}
      {\bf 1}\{J\subseteq\mathcal{J}_t(i)\}
      \pi_t(\mathcal{R}|i,J)\enspace.
\]

Uniform sampling from (\ref{eq:control_set}) is a special case of nested case-control sampling \citep{bgl-mascdcphm-95}, in particular it is $\pi_t(\mathcal{R}|i,J)=\pi_t(\mathcal{R}|i,J')$ for all $J,J'\in\mathcal{R}$, and by \citet{bgl-mascdcphm-95} we can estimate parameters $\hat{\beta}$ by maximizing the log partial likelihood function
  \begin{equation}
    \label{eq:likelihood_rhem_sampled}
    \log \tilde{L}_t(\beta)=\sum_{t_m\leq t}\left(
    \beta^{\rm T}x_{t_m}(i_m,J_m)
    -
    \log\left[
      \sum_{J\in \tilde{\mathcal{R}}_{t_m}(\mathcal{J}_{t_m}(i_m),J_m,k)}\exp\left\{
        \beta^{\rm T}x_{t_m}(i_m,J)
        \right\}
      \right]
    \right)\enspace.
  \end{equation}
  \citet{bgl-mascdcphm-95} prove conditions under which the MPLE is a consistent estimator. Estimating parameters based on the partial likelihood (\ref{eq:likelihood_rhem_sampled}) uses only information about the selection of receiver sets $J_m$ out of all \emph{sampled} sets of receivers $J\in \tilde{\mathcal{R}}_{t_m}(\mathcal{J}_{t_m}(i_m),J_m,k)$. On the other hand, it does not consider information contained in the event time $t_m$, sender $i_m$, and receiver set size $|J_m|$, which are absorbed in the baseline rate $\overline{\lambda}_t(i,|J|)$. Therefore, including covariates that are only functions of event times, senders, or size of the receiver sets would lead to non-identifiable parameters.

\subsection{Hyperedge covariates}
\label{sec:covar_rhem}

Since hyperedge covariates $x_t(i,J)$ are defined on the entire set of receivers, a large number of structurally different covariates is possible. While our list is certainly far from exhaustive, we define in the following a collection of hyperedge covariates that are practically relevant in the empirical analysis of multicast interaction networks. We define two types of covariates: ``attribute effects'', dependent on actor-level attributes and ``network effects,'' dependent on the history of the process, that is, on sequences of previously observed events.

\subsubsection{Covariates dependent on actor-level attributes}
\label{sec:covar_actor-level}

Suppose that available data include information on one or several numeric actor-level attributes $z\colon \mathcal{A}\to\mathbb{R}$. Then, in general, RHEM covariates $x_t(i,J)$ based on such actor (``node-specific'') characteristics may be obtained either by functions of the values in the receiver set such as $\textup{mean}\{z(j)\,;\;j\in J\}$, or by functions of the receivers' values in relation to the sender's value $z(i)$, such as $\textup{mean}\{|z(j)-z(i)|\,;\;j\in J\}$. Covariates that are only functions of the sender $i$ would lead to a non-identifiable parameter, since their effect would be absorbed by the baseline rate $\overline{\lambda}_t(i,|J|)$. 

Concrete examples of covariates are discussed below, where we also discuss the possibility of defining covariates based on categorical attributes. In this article we consider only time-invariant attributes. However, the covariates discussed below extend directly to time-varying actor-level attributes by considering values at the given time $t$.

\paragraph{Receiver set average} of attribute $z$, abbreviated as $rec\_avg\_z$, is defined to be
\[
rec\_avg\_z_t(i,J)=\sum_{j\in J}z(j)/|J|\enspace.
\]
For example, in the empirical analysis we apply this definition to four binary attributes \emph{female}, \emph{senior}, \emph{trade}, and \emph{legal}. The receiver set average, thus, gives the proportion of receivers that are females, seniors, members of the ``trade'' department, or members of the ``legal'' department, respectively. The definition also applies to non-binary numeric attributes.

\paragraph{Sender-receiver heterophily.} We assess heterophily between senders and receivers by averaging the absolute difference between the attribute value of the sender $i$ and the values of the receivers $j\in J$:
\[
send\_rec\_diff\_z_t(i,J)=\sum_{j\in J}|z(i)-z(j)|/|J|\enspace.
\]
For example, for the attribute \emph{female} we obtain the fraction of receivers having the opposite gender than the sender.
If senders have a tendency to interact with similar receivers, then a higher value of sender-receiver homophily would make an interaction on the hyperedge $(i,J)$ less likely. Thus, in presence of homophily we would expect a negative parameter -- a positive parameter would point to heterophily.

In the case of a categorical attribute $z$, such as the attribute ``department'', we assess sender-receiver heterophily via the fraction of receivers that have a different value than the sender:
\[
send\_rec\_diff\_z_t(i,J)=\sum_{j\in J}{\bf 1}\{z(i)\neq z(j)\}/|J|\enspace.
\]

\paragraph{Receiver set heterophily.} We assess heterophily in the receiver set by averaging the absolute difference of the attribute value between receiver pairs:
\[
rec\_set\_diff\_z_t(i,J)=\sum_{\{j,j'\}\in {J \choose 2}}\frac{|z(j')-z(j)|}{{|J| \choose 2}}\enspace.
\]
If senders have a tendency to interact with homogeneous receiver sets (irrespective of their own attribute value), then a higher value of the covariate capturing receiver set homophily would make an interaction less likely. Thus, we would expect a negative parameter in the presence of homophily; a positive parameter would point to heterophily.

For a categorical attribute $z$, we assess receiver set heterophily via the fraction of receiver pairs that have a different attribute value:
\[
rec\_set\_diff\_z_t(i,J)=\sum_{\{j,j'\}\in {J \choose 2}}\frac{{\bf 1}\{z(j')\neq z(j)\}}{{|J| \choose 2}}\enspace.
\]

Note that receiver set heterophily is structurally different from sender-receiver heterophily \citep{snijders2019beyond}. Theoretically it might be the case that senders have neither a preference, nor a reluctance to send messages to, say, receivers in their own department. Yet, in the same data it is possible that typical receiver sets are mostly composed of members from one department. In such a hypothetical scenario, we would not find evidence for sender-receiver homophily but we would find evidence for receiver set homophily.

\subsubsection{Network effects}
\label{sec:covar_network}

In addition to covariates based on actor-level attributes, we define several covariates expressing dependence of the intensity $\lambda_t(i,J)$ on the history of the process. Given an observed sequence of events generated by polyadic interaction  $E=\{(t_1,i_1,J_1),\dots,(t_n,i_n,J_n)\}$, the value of these covariates at time $t$ is computed as a function of earlier events $E_{<t}=\{(t_m,i_m,J_m)\in E\,;\;t_m<t\}$.

Similar to previous work on REM and RHEM \citep{brandes2009networks,lerner2013modeling,amati2019some,lerner2021dynamic}, we let the effect of past events decay over time. More precisely, for a given half life period $T_{1/2}>0$, the weight of a past event $(t_m,i_m,J_m)$ at current time $t>t_m$ is defined to be $w(t-t_m)=\exp\left(-(t-t_m)\frac{\log 2}{T_{1/2}}\right)$. While alternatives exist to exponential decay of past events, \citet{schecter2021power} found it to be generally adequate. The objective of our paper is to clarify the benefit of hyperedge covariates in comparison to dyadic covariates. We consider issues related to the effect of elapsed time of past events as orthogonal to the  objective of the current paper. Hyperedge covariates and dyadic covariates considered in the empirical part of this paper are defined with the same decay mechanism and the same half life period.

\paragraph{Exact repetition.} An empirically plausible effect would capture relational ``inertia,'' or the tendency of actors to continue to do what they did in the past. In fact, in many cases it may be important to verify the presence of more complex mechanisms over and above the simple repetition of past behavior. In polyadic interaction networks this inertial behavioral tendency would lead to future events that repeat the sender and the entire receiver set of past events. Such an effect can be captured by the covariate:
\[
exact\_repetition_t(i,J)=\sum_{(t_m,i_m,J_m)\in E_{<t}}w(t-t_m)\cdot{\bf 1}(i_m=i\wedge J_m=J)\enspace.
\]
Thus, we sum the current weight over all previous events that have the same sender and exactly the same receiver set. Exact repetition captures effects in which the same sender repeatedly addresses the same receivers, for instance, email communication with fixed receiver lists (``mailing lists'').

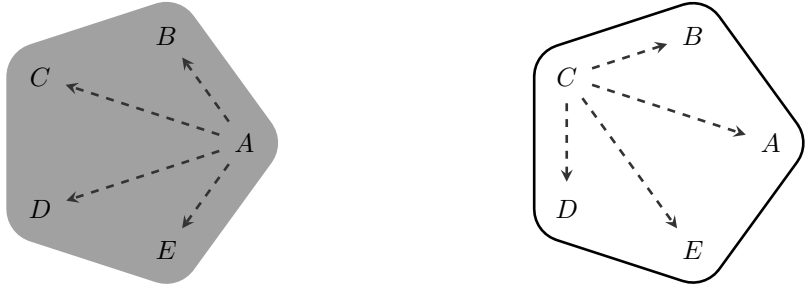
\begin{figure}
  \centering
  \makebox{%
    \begin{tikzpicture}[scale=1,>=stealth]
      \node at (-4,0) (left) {%
      \begin{tikzpicture}[scale=1]
        \tikzstyle{actor}=[circle,minimum size=1mm];
        \node[actor] at (0:1.5) (a) {$A$};
        \node[actor] at (72:1.5) (b) {$B$};
        \node[actor] at (144:1.5) (c) {$C$};
        \node[actor] at (216:1.5) (d) {$D$};
        \node[actor] at (288:1.5) (e) {$E$};
        
        \draw[color=darkgrey,line width=1pt,dashed,->] (a) to (b);
        \draw[color=darkgrey,line width=1pt,dashed,->] (a) to (c);
        \draw[color=darkgrey,line width=1pt,dashed,->] (a) to (d);
        \draw[color=darkgrey,line width=1pt,dashed,->] (a) to (e);

        \begin{pgfonlayer}{background}
          \foreach \nodename in {a,b,c,d,e} {
            \coordinate (\nodename') at (\nodename);
          }
          \path[fill=grey1,draw=grey1,line width=0.9cm, line cap=round, line join=round] 
          (a') to (b') to (c') to (d') to (e') to (a') -- cycle;
        \end{pgfonlayer}
      \end{tikzpicture}};
      \node at (3,0) (right) {%
      \begin{tikzpicture}[scale=1]
        \tikzstyle{actor}=[circle,minimum size=1mm];
        \node[actor] at (0:1.5) (a) {$A$};
        \node[actor] at (72:1.5) (b) {$B$};
        \node[actor] at (144:1.5) (c) {$C$};
        \node[actor] at (216:1.5) (d) {$D$};
        \node[actor] at (288:1.5) (e) {$E$};
        
        \draw[color=darkgrey,line width=1pt,dashed,->] (c) to (b);
        \draw[color=darkgrey,line width=1pt,dashed,->] (c) to (a);
        \draw[color=darkgrey,line width=1pt,dashed,->] (c) to (d);
        \draw[color=darkgrey,line width=1pt,dashed,->] (c) to (e);

        \begin{pgfonlayer}{background}
          \foreach \nodename in {a,b,c,d,e} {
            \coordinate (\nodename') at (\nodename);
          }
          \path[fill=black,draw=black,line width=0.9cm, line cap=round, line join=round] 
          (a') to (b') to (c') to (d') to (e') to (a') -- cycle;
          \path[fill=white,draw=white,line width=0.83cm, line cap=round, line join=round] 
          (a') to (b') to (c') to (d') to (e') to (a') -- cycle;
        \end{pgfonlayer}
      \end{tikzpicture}};
    \end{tikzpicture}
  }
  \caption{\footnotesize \label{fig:repetition} Stylized example illustrating unordered repetition. \textit{Left}: history of a past event $e_1=(t_1,A,\{B,C,D,E\})$ displayed as a gray-shaded area; dashed lines connect the sender to the receivers. \textit{Right}: a candidate hyperedge $h=(C,\{A,B,D,E\})$ for a future hyperevent. The past event $e_1$ increases the value of unordered repetition on $h$ at time $t>t_1$. In communication networks, unordered repetition could point to turn-taking among a stable set of conversation participants.}
\end{figure}

\paragraph{Unordered repetition.} Besides exact repetition, there can be situations of interaction within a stable group of actors with turn-taking among the senders \citep{gibson2005taking}. Such an effect can be captured by the covariate
\[
unordered\_repetition_t(i,J)=\sum_{(t_m,i_m,J_m)\in E_{<t}}w(t-t_m)\cdot{\bf 1}(\{i_m\}\cup J_m=\{i\}\cup J)\enspace.
\]
In contrast to (ordered) exact repetition, the unordered repetition effect allows that different actors take on the role of sender -- as long as the union of the sender with the receiver set remains constant. A typical example of this behavior is email communication using a ``reply-to-all'' functionality: a receiver of a previous message sends a message to the previous sender and to all other receivers. See Fig.~\ref{fig:repetition} for an illustration of unordered repetition. In contrast to unordered repetition, exact repetition would require that the future event in Fig.~\ref{fig:repetition} is exactly on the hyperedge $(A,\{B,C,D,E\})$, that is, the same sender $A$ sends another interaction to the same receiver set $\{B,C,D,E\}$.

\paragraph{Partial receiver set repetition.} The two (exact and unordered) repetition covariates defined above still give an incomplete picture of stability in multicast interaction events, since they require that sets of actors involved in past and current events have to be identical. A possible event that mostly, but not exactly, repeats the receiver set of a past event -- for instance, if some new receivers are added and/or if some previous receivers are removed -- is treated identically to a possible event with a completely disjoint receiver set. To quantify partial repetition, we define a parametric family of covariates capturing to what extent subsets of a possible receiver set have jointly received past events.

To shorten notation, we define the \emph{hyperedge indegree} of a set of receivers $J'\subseteq\mathcal{J}$ by considering past events that have been jointly received by all members of $J'$ -- possibly together with varying other receivers outside of $J'$:
\[
hy\_deg^{(in)}_t(J')=\sum_{(t_m,i_m,J_m)\in E_{<t}}w(t-t_m)\cdot{\bf 1}(J'\subseteq J_m)\enspace.
\]

Partial receiver set repetition (or subset repetition among the receivers) is parametrized by a positive integer $p$, giving the cardinality of the subsets that repeatedly receive joint messages:
\[
rec\_sub\_rep^{(p)}_t(i,J)=\sum_{J'\in {J \choose p}}\frac{hy\_deg^{(in)}_t(J')}{{|J| \choose p}}\enspace.
\]
For $p=1$ we obtain the average indegree of individual receivers $j\in J$ by considering past interactions received by $j$, downweighted by the elapsed time. For $p=2$ we consider past interactions that have been jointly received by pairs $\{j,j'\}\subseteq J$, and so on.

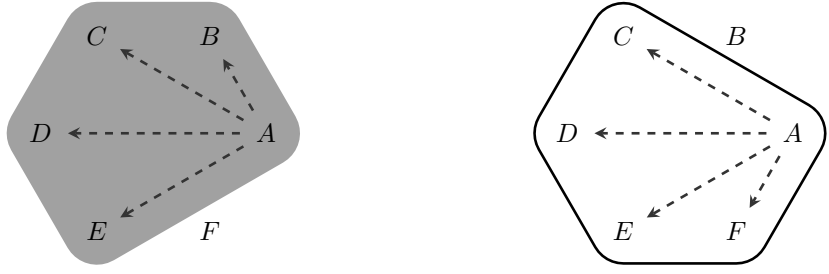
\begin{figure}
  \centering
  \makebox{%
    \begin{tikzpicture}[scale=1,>=stealth]
      \node at (-4,0) (left) {%
      \begin{tikzpicture}[scale=1]
        \tikzstyle{actor}=[circle,minimum size=1mm];
        \node[actor] at (0:1.5) (a) {$A$};
        \node[actor] at (60:1.5) (b) {$B$};
        \node[actor] at (120:1.5) (c) {$C$};
        \node[actor] at (180:1.5) (d) {$D$};
        \node[actor] at (240:1.5) (e) {$E$};
        \node[actor] at (300:1.5) (f) {$F$};
        
        \draw[color=darkgrey,line width=1pt,dashed,->] (a) to (b);
        \draw[color=darkgrey,line width=1pt,dashed,->] (a) to (c);
        \draw[color=darkgrey,line width=1pt,dashed,->] (a) to (d);
        \draw[color=darkgrey,line width=1pt,dashed,->] (a) to (e);

        \begin{pgfonlayer}{background}
          \foreach \nodename in {a,b,c,d,e,f} {
            \coordinate (\nodename') at (\nodename);
          }
          \path[fill=grey1,draw=grey1,line width=0.9cm, line cap=round, line join=round] 
          (a') to (b') to (c') to (d') to (e') to (a') -- cycle;
        \end{pgfonlayer}
      \end{tikzpicture}};
      \node at (3,0) (right) {%
      \begin{tikzpicture}[scale=1]
        \tikzstyle{actor}=[circle,minimum size=1mm];
        \node[actor] at (0:1.5) (a) {$A$};
        \node[actor] at (60:1.5) (b) {$B$};
        \node[actor] at (120:1.5) (c) {$C$};
        \node[actor] at (180:1.5) (d) {$D$};
        \node[actor] at (240:1.5) (e) {$E$};
        \node[actor] at (300:1.5) (f) {$F$};
        
        \draw[color=darkgrey,line width=1pt,dashed,->] (a) to (f);
        \draw[color=darkgrey,line width=1pt,dashed,->] (a) to (c);
        \draw[color=darkgrey,line width=1pt,dashed,->] (a) to (d);
        \draw[color=darkgrey,line width=1pt,dashed,->] (a) to (e);

        \begin{pgfonlayer}{background}
          \foreach \nodename in {a,b,c,d,e,f} {
            \coordinate (\nodename') at (\nodename);
          }
          \path[fill=black,draw=black,line width=0.9cm, line cap=round, line join=round] 
          (a') to (c') to (d') to (e') to (f') to (a') -- cycle;
          \path[fill=white,draw=white,line width=0.83cm, line cap=round, line join=round] 
          (a') to (c') to (d') to (e') to (f') to (a') -- cycle;
        \end{pgfonlayer}
      \end{tikzpicture}};
    \end{tikzpicture}
  }
  \caption{\footnotesize \label{fig:partial_repetition} Stylized example illustrating (sender-specific) partial receiver set repetition. \textit{Left}: history of a past event $e_1=(t_1,A,\{B,C,D,E\})$. \textit{Right}: a candidate hyperedge $h=(A,\{C,D,E,F\})$ for a future hyperevent. Among the four receivers in $h$, three have individually received the past event $e_1$. Among the six pairs of receivers in $h$, three have jointly received the past event $e_1$. Among the four triples of receivers in $h$, one has jointly received the past event $e_1$. The past event $e_1$ increases the value of sender-specific partial receiver set repetition of order $p=1,2,3$ on $h$ at $t>t_1$. If the sender of $h$ was another actor $G$, instead of $A$, then the past event would still increase the value of partial receiver set repetition, but it would not increase the value of the sender-specific variant.}
\end{figure}

Partial receiver set repetition -- and the related sender-specific partial receiver set repetition, defined below -- are illustrated in Fig.~\ref{fig:partial_repetition}. In the notation of that figure, if we ignore the decay over time and if $e_1$ is the entire history, we get for a point in time $t>t_1$ the following values:
\begin{eqnarray*}
rec\_sub\_rep^{(1)}_t(h)&=&3/4\\
rec\_sub\_rep^{(2)}_t(h)&=&3/6\\
rec\_sub\_rep^{(3)}_t(h)&=&1/4\enspace.
\end{eqnarray*}
Partial receiver set repetition of order $p>3$ is zero in this example.

\paragraph{Sender-specific partial receiver set repetition.} Partial receiver set repetition defined above does not consider whether past interactions jointly received by $J'\subseteq J$ originated from the same sender $i$. Thus, these covariates capture partial repetition of receivers by any sender. To consider only partial receiver set repetition by the same sender, we first define the \emph{sender-specific hyperedge degree} by :
\[
hy\_deg_t(i,J')=\sum_{(t_m,i_m,J_m)\in E_{<t}}w(t-t_m)\cdot{\bf 1}(i=i_m\wedge J'\subseteq J_m)\enspace.
\]

Sender-specific partial receiver set repetition (or sender-specific subset repetition in the receiver set) is parametrized by a positive integer $p$, giving the cardinality of the subset that repeatedly receives joint messages from the given sender $i$.
\[
send\_rec\_sub\_rep^{(p)}_t(i,J)=\sum_{J'\in {J \choose p}}\frac{hy\_deg_t(i,J')}{{|J| \choose p}}\enspace.
\]
For $p=1$ we obtain the average weight of past interactions that individual actors $j\in J$ received from the given sender $i$, where the average is taken over all those receivers. For $p=2$ we consider past interactions that have been sent by $i$ and that have been jointly received by pairs $\{j,j'\}\subseteq J$, and so on.

In the example illustrated in Fig.~\ref{fig:partial_repetition} we obtain the same values for sender-specific partial receiver set repetition as for partial receiver set repetition, since the candidate hyperedge for a future event $h=(A,\{C,D,E,F\})$ has the same sender as the past event $e_1$. However, another candidate hyperedge $h'=(G,\{C,D,E,F\})$ would have the same value as $h$ in the partial receiver set repetition covariates but would get the value zero in the sender-specific variants.

(Sender-specific) partial receiver set repetition of order $p\geq 2$ may induce a clustering in the set of actors, revealing subsets of actors that are likely to jointly receive the same interactions. Partial receiver set repetition implies a ``global'' clustering applying to all interactions, irrespective of their sender. The sender-specific variants allow for different clusterings of receivers that vary with the sender.

\begin{figure}
  \centering
  \makebox{%
    \begin{tikzpicture}[scale=1,>=stealth]
      \node at (-4,0) (left) {%
      \begin{tikzpicture}[scale=1]
        \tikzstyle{actor}=[circle,minimum size=1mm];
        \node[actor] at (0:1.5) (a) {$A$};
        \node[actor] at (60:1.5) (b) {$B$};
        \node[actor] at (120:1.5) (c) {$C$};
        \node[actor] at (180:1.5) (d) {$D$};
        \node[actor] at (240:1.5) (e) {$E$};
        \node[actor] at (300:1.5) (f) {$F$};
        
        \draw[color=darkgrey,line width=1pt,dashed,->] (a) to (c);
        \draw[color=darkgrey,line width=1pt,dashed,->] (a) to (d);
        \draw[color=darkgrey,line width=1pt,dashed,->] (a) to (e);

        \begin{pgfonlayer}{background}
          \foreach \nodename in {a,b,c,d,e,f} {
            \coordinate (\nodename') at (\nodename);
          }
          \path[fill=grey1,draw=grey1,line width=0.9cm, line cap=round, line join=round] 
          (a') to (c') to (d') to (e') to (a') -- cycle;
        \end{pgfonlayer}
      \end{tikzpicture}};
      \node at (3,0) (right) {%
      \begin{tikzpicture}[scale=1]
        \tikzstyle{actor}=[circle,minimum size=1mm];
        \node[actor] at (0:1.5) (a) {$A$};
        \node[actor] at (60:1.5) (b) {$B$};
        \node[actor] at (120:1.5) (c) {$C$};
        \node[actor] at (180:1.5) (d) {$D$};
        \node[actor] at (240:1.5) (e) {$E$};
        \node[actor] at (300:1.5) (f) {$F$};
        
        \draw[color=darkgrey,line width=1pt,dashed,->] (f) to (a);
        \draw[color=darkgrey,line width=1pt,dashed,->] (f) to (c);
        \draw[color=darkgrey,line width=1pt,dashed,->] (f) to (d);
        \draw[color=darkgrey,line width=1pt,dashed,->] (f) to (b);

        \begin{pgfonlayer}{background}
          \foreach \nodename in {a,b,c,d,e,f} {
            \coordinate (\nodename') at (\nodename);
          }
          \path[fill=black,draw=black,line width=0.9cm, line cap=round, line join=round] 
          (a') to (b') to (c') to (d') to (f') to (a') -- cycle;
          \path[fill=white,draw=white,line width=0.83cm, line cap=round, line join=round] 
          (a') to (b') to (c') to (d') to (f') to (a') -- cycle;
        \end{pgfonlayer}
      \end{tikzpicture}};
    \end{tikzpicture}
  }
  \caption{\footnotesize \label{fig:interaction_among_receivers} Stylized example illustrating the covariate ``interaction among receivers''. \textit{Left}: history of a past event $e_1=(t_1,A,\{C,D,E\})$. \textit{Right}: a candidate hyperedge $h=(F,\{A,B,C,D\})$ for a future hyperevent. The sender $F$ of $h$ sends an interaction to the sender $A$ of the past event $e_1$ and to two of its receivers ($C$ and $D$). The past event $e_1$ increases the value of the covariate interaction among receivers on $h$ at $t>t_1$ for $p=1,2$. For $p>2$ that covariate is zero since there are no three previous receivers that receive an interaction together with the previous sender.} 
\end{figure}
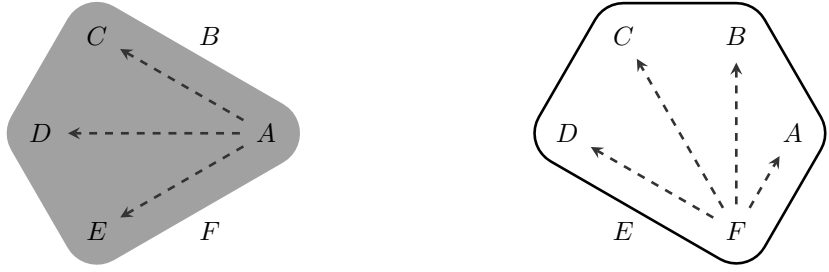

\paragraph{Past interaction among receivers.} Yet another network effect in polyadic social interaction network arises if actors send interactions to a sender of a past interaction together with a subset of the receivers of that past interaction. This is, for instance, a frequent pattern in scientific citation networks where a paper $P$ cites another paper $P'$ and some of the references of $P'$. This pattern can be captured by the following family of covariates, parametrized by a positive integer $p$, giving the number of the repeated receivers of the previous event:
\[
interact\_rec^{(p)}_t(i,J)=\sum_{j\in J,\,J'\in {J\setminus\{j\} \choose p}}\frac{hy\_deg_t(j,J')}{|J|\cdot {|J|-1 \choose p}}\enspace.
\]
Interaction among receivers is illustrated in Fig.~\ref{fig:interaction_among_receivers}. In the notation of that figure, if we ignore the decay over time and if $e_1$ is the entire history, we get for a point in time $t>t_1$ the following values:
\begin{eqnarray*}
interact\_rec^{(1)}_t(h)&=&2/(4\cdot3)\\
interact\_rec^{(2)}_t(h)&=&1/(4\cdot3)\enspace.
\end{eqnarray*}
Interaction among receivers of order $p>2$ is zero in this example.

\paragraph{Reciprocation and out-in popularity.} Actors often have the tendency to reply to interactions they received in the past. Reciprocation is captured by considering interactions that the sender $i$ of the current interaction has received from actors $j\in J$:
\[
recip_t(i,J)=\sum_{j\in J}hy\_deg_t(j,\{i\})/|J|\enspace.
\]

\begin{figure}
  \centering
  \makebox{%
    \begin{tikzpicture}[scale=1,>=stealth]
      \node at (-4,0) (left) {%
      \begin{tikzpicture}[scale=1]
        \tikzstyle{actor}=[circle,minimum size=1mm];
        \node[actor] at (0:1.5) (a) {$A$};
        \node[actor] at (60:1.5) (b) {$B$};
        \node[actor] at (120:1.5) (c) {$C$};
        \node[actor] at (180:1.5) (d) {$D$};
        \node[actor] at (240:1.5) (e) {$E$};
        \node[actor] at (300:1.5) (f) {$F$};
        
        \draw[color=darkgrey,line width=1pt,dashed,->] (a) to (f);
        \draw[color=darkgrey,line width=1pt,dashed,->] (a) to (d);
        \draw[color=darkgrey,line width=1pt,dashed,->] (a) to (e);
        \draw[color=darkgrey,line width=1pt,dashed,->] (b) to (c);
        \draw[color=darkgrey,line width=1pt,dashed,->] (b) to (a);

        \begin{pgfonlayer}{background}
          \foreach \nodename in {a,b,c,d,e,f} {
            \coordinate (\nodename') at (\nodename);
          }
          \path[fill=grey1,draw=grey1,line width=0.9cm, line cap=round, line join=round] 
          (a') to (d') to (e') to (f') to (a') -- cycle;
          \path[fill=grey3,draw=grey3,line width=0.7cm, line cap=round, line join=round] 
          (a') to (b') to (c') to (a') -- cycle;
        \end{pgfonlayer}
      \end{tikzpicture}};
      \node at (3,0) (right) {%
      \begin{tikzpicture}[scale=1]
        \tikzstyle{actor}=[circle,minimum size=1mm];
        \node[actor] at (0:1.5) (a) {$A$};
        \node[actor] at (60:1.5) (b) {$B$};
        \node[actor] at (120:1.5) (c) {$C$};
        \node[actor] at (180:1.5) (d) {$D$};
        \node[actor] at (240:1.5) (e) {$E$};
        \node[actor] at (300:1.5) (f) {$F$};
        
        \draw[color=darkgrey,line width=1pt,dashed,->] (d) to (a);
        \draw[color=darkgrey,line width=1pt,dashed,->] (d) to (b);
        \draw[color=darkgrey,line width=1pt,dashed,->] (d) to (c);

        \begin{pgfonlayer}{background}
          \foreach \nodename in {a,b,c,d,e,f} {
            \coordinate (\nodename') at (\nodename);
          }
          \path[fill=black,draw=black,line width=0.9cm, line cap=round, line join=round] 
          (a') to (b') to (c') to (d') to (a') -- cycle;
          \path[fill=white,draw=white,line width=0.83cm, line cap=round, line join=round] 
          (a') to (b') to (c') to (d') to (a') -- cycle;
        \end{pgfonlayer}
      \end{tikzpicture}};
    \end{tikzpicture}
  }
  \caption{\footnotesize \label{fig:reciprocation} Stylized example illustrating reciprocation and out-in popularity. \textit{Left}: history of two past events $e_1=(t_1,A,\{D,E,F\})$ and $e_2=(t_2,B,\{A,C\})$. \textit{Right}: a candidate hyperedge $h=(D,\{A,B,C\})$ for a future hyperevent. The past event $e_1$ increases the value of the reciprocation covariate on $h$ at $t>t_1$, since $e_1$ has been sent by $A$, a receiver of $h$, among others to $D$, the sender of $h$. The past event $e_2$ does not increase reciprocation on $h$ -- but it increases out-in popularity on $h$, since $e_2$ has been send by $B$, a receiver of $h$.} 
\end{figure}
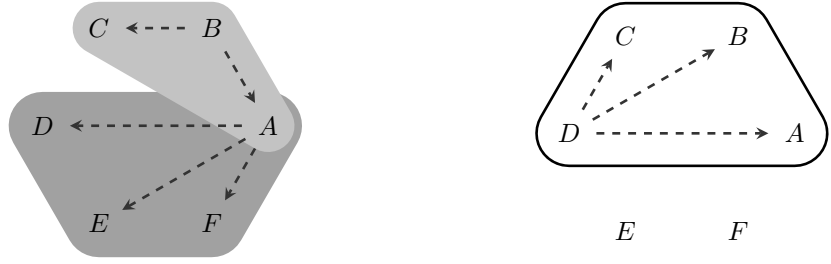

The effect \emph{out-in popularity}, arises when senders of past interactions receive future interactions from any actor in the network -- not necessarily from the receivers of the past interactions. We define the \emph{outdegree} of an actor $i'\in\mathcal{A}$ by considering past events that have been send by $i'$. (Note that in contrast to the hyperedge indegree, the outdegree cannot be defined for a set of more than one actor, since every interaction has only one sender.)
\[
deg^{(out)}_t(i')=\sum_{(t_m,i_m,J_m)\in E_{<t}}w(t-t_m)\cdot{\bf 1}(i'=i_m)\enspace.
\]
Out-in popularity is defined by
\[
out\_in\_pop_t(i,J)=\sum_{j\in J}deg^{(out)}_t(j)/|J|\enspace.
\]
Reciprocation and out-in popularity are illustrated in Fig.~\ref{fig:reciprocation}. In the notation of that figure, if we ignore the decay over time and if $e_1,e_2$ is the entire history, we get for a point in time $t>t_1,t_2$ the following values:
\begin{eqnarray*}
recip_t(h)&=&1/3\\
out\_in\_pop_t(h)&=&2/3\enspace.
\end{eqnarray*}

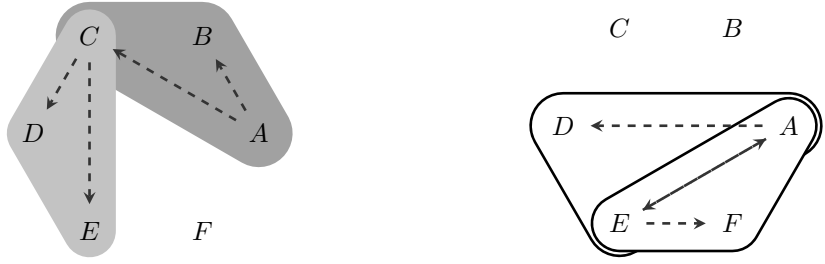
\begin{figure}
  \centering
  \makebox{%
    \begin{tikzpicture}[scale=1,>=stealth]
      \node at (-4,0) (left) {%
      \begin{tikzpicture}[scale=1]
        \tikzstyle{actor}=[circle,minimum size=1mm];
        \node[actor] at (0:1.5) (a) {$A$};
        \node[actor] at (60:1.5) (b) {$B$};
        \node[actor] at (120:1.5) (c) {$C$};
        \node[actor] at (180:1.5) (d) {$D$};
        \node[actor] at (240:1.5) (e) {$E$};
        \node[actor] at (300:1.5) (f) {$F$};
        
        \draw[color=darkgrey,line width=1pt,dashed,->] (a) to (b);
        \draw[color=darkgrey,line width=1pt,dashed,->] (a) to (c);
        \draw[color=darkgrey,line width=1pt,dashed,->] (c) to (d);
        \draw[color=darkgrey,line width=1pt,dashed,->] (c) to (e);

        \begin{pgfonlayer}{background}
          \foreach \nodename in {a,b,c,d,e,f} {
            \coordinate (\nodename') at (\nodename);
          }
          \path[fill=grey1,draw=grey1,line width=0.9cm, line cap=round, line join=round] 
          (a') to (b') to (c') to (a') -- cycle;
          \path[fill=grey3,draw=grey3,line width=0.7cm, line cap=round, line join=round] 
          (c') to (d') to (e') to (c') -- cycle;
        \end{pgfonlayer}
      \end{tikzpicture}};
      \node at (3,0) (right) {%
      \begin{tikzpicture}[scale=1]
        \tikzstyle{actor}=[circle,minimum size=1mm];
        \node[actor] at (0:1.5) (a) {$A$};
        \node[actor] at (60:1.5) (b) {$B$};
        \node[actor] at (120:1.5) (c) {$C$};
        \node[actor] at (180:1.5) (d) {$D$};
        \node[actor] at (240:1.5) (e) {$E$};
        \node[actor] at (300:1.5) (f) {$F$};
        
        \draw[color=darkgrey,line width=1pt,dashed,->] (a) to (d);
        \draw[color=darkgrey,line width=1pt,dashed,->] (a) to (e);
        \draw[color=darkgrey,line width=1pt,dashed,->] (e) to (a);
        \draw[color=darkgrey,line width=1pt,dashed,->] (e) to (f);

        \begin{pgfonlayer}{background}
          \foreach \nodename in {a,b,c,d,e,f} {
            \coordinate (\nodename') at (\nodename);
          }
          \path[fill=black,draw=black,line width=0.9cm, line cap=round, line join=round] 
          (a') to (d') to (e') to (a') -- cycle;
          \path[fill=white,draw=white,line width=0.83cm, line cap=round, line join=round] 
          (a') to (d') to (e') to (a') -- cycle;
          \path[fill=black,draw=black,line width=0.77cm, line cap=round, line join=round] 
          (a') to (e') to (f') to (a') -- cycle;
          \path[fill=white,draw=white,line width=0.7cm, line cap=round, line join=round] 
          (a') to (e') to (f') to (a') -- cycle;
        \end{pgfonlayer}
      \end{tikzpicture}};
    \end{tikzpicture}
  }
  \caption{\footnotesize \label{fig:closure_1} Stylized example illustrating transitive closure and cyclic closure. \textit{Left}: history of two past events $e_1=(t_1,A,\{B,C\})$ and $e_2=(t_2,C,\{D,E\})$. \textit{Right}: two candidate hyperedges $h=(A,\{D,E\})$ and $h'=(E,\{A,F\})$ for future hyperevents. The past events $e_1,e_2$ increase the value of transitive closure on $h$ at $t>t_1,t_2$, since $h$ transitively closes two paths: from $A$ over $C$ to $D$ and from $A$ over $C$ to $E$. The past events $e_1,e_2$ increase the value of cyclic closure on $h'$ at $t>t_1,t_2$, since $h'$ closes a cycle from $A$ to $C$ to $E$ to $A$.} 
\end{figure}

\paragraph{Triadic effects.} Interactions might further depend on past interactions that the sender and the receivers had with common third actors. By varying the directions of the past interactions to or from the sender and the receivers we obtain four variants of triadic closure, denoted by transitive closure, cyclic closure, incoming balance, and outgoing balance. In the summations below, $a$ iterates over all actors $\mathcal{A}\setminus\{i,j\}$.
\begin{eqnarray*}
transitive\_closure_t(i,J)&=&\sum_{j\in J,\,a\neq i,j}\frac{\min\left\{hy\_deg_t(i,\{a\}),hy\_deg_t(a,\{j\})\right\}}{|J|}\\
cyclic\_closure_t(i,J)&=&\sum_{j\in J,\,a\neq i,j}\frac{\min\left\{hy\_deg_t(a,\{i\}),hy\_deg_t(j,\{a\})\right\}}{|J|}\\
in\_balance_t(i,J)&=&\sum_{j\in J,\,a\neq i,j}\frac{\min\left\{hy\_deg_t(a,\{i\}),hy\_deg_t(a,\{j\})\right\}}{|J|}\\
out\_balance_t(i,J)&=&\sum_{j\in J,\,a\neq i,j}\frac{\min\left\{hy\_deg_t(i,\{a\}),hy\_deg_t(j,\{a\})\right\}}{|J|}\enspace.\\
\end{eqnarray*}
Transitive closure and cyclic closure are illustrated in Fig.~\ref{fig:closure_1}. In the notation of that figure, if we ignore the decay over time and if $e_1,e_2$ is the entire history, for a point in time $t>t_1,t_2$ we obtain the following values:
\begin{eqnarray*}
transitive\_closure_t(h)&=&2/2\\
cyclic\_closure_t(h)&=&1/2\enspace.
\end{eqnarray*}

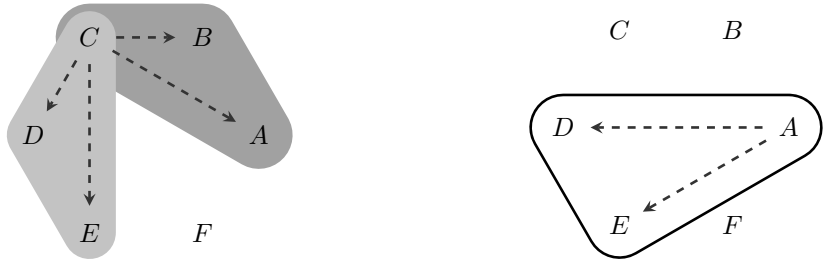
\begin{figure}
  \centering
  \makebox{%
    \begin{tikzpicture}[scale=1,>=stealth]
      \node at (-4,0) (left) {%
      \begin{tikzpicture}[scale=1]
        \tikzstyle{actor}=[circle,minimum size=1mm];
        \node[actor] at (0:1.5) (a) {$A$};
        \node[actor] at (60:1.5) (b) {$B$};
        \node[actor] at (120:1.5) (c) {$C$};
        \node[actor] at (180:1.5) (d) {$D$};
        \node[actor] at (240:1.5) (e) {$E$};
        \node[actor] at (300:1.5) (f) {$F$};
        
        \draw[color=darkgrey,line width=1pt,dashed,->] (c) to (b);
        \draw[color=darkgrey,line width=1pt,dashed,->] (c) to (a);
        \draw[color=darkgrey,line width=1pt,dashed,->] (c) to (d);
        \draw[color=darkgrey,line width=1pt,dashed,->] (c) to (e);

        \begin{pgfonlayer}{background}
          \foreach \nodename in {a,b,c,d,e,f} {
            \coordinate (\nodename') at (\nodename);
          }
          \path[fill=grey1,draw=grey1,line width=0.9cm, line cap=round, line join=round] 
          (a') to (b') to (c') to (a') -- cycle;
          \path[fill=grey3,draw=grey3,line width=0.7cm, line cap=round, line join=round] 
          (c') to (d') to (e') to (c') -- cycle;
        \end{pgfonlayer}
      \end{tikzpicture}};
      \node at (3,0) (right) {%
      \begin{tikzpicture}[scale=1]
        \tikzstyle{actor}=[circle,minimum size=1mm];
        \node[actor] at (0:1.5) (a) {$A$};
        \node[actor] at (60:1.5) (b) {$B$};
        \node[actor] at (120:1.5) (c) {$C$};
        \node[actor] at (180:1.5) (d) {$D$};
        \node[actor] at (240:1.5) (e) {$E$};
        \node[actor] at (300:1.5) (f) {$F$};
        
        \draw[color=darkgrey,line width=1pt,dashed,->] (a) to (d);
        \draw[color=darkgrey,line width=1pt,dashed,->] (a) to (e);

        \begin{pgfonlayer}{background}
          \foreach \nodename in {a,b,c,d,e,f} {
            \coordinate (\nodename') at (\nodename);
          }
          \path[fill=black,draw=black,line width=0.9cm, line cap=round, line join=round] 
          (a') to (d') to (e') to (a') -- cycle;
          \path[fill=white,draw=white,line width=0.83cm, line cap=round, line join=round] 
          (a') to (d') to (e') to (a') -- cycle;
        \end{pgfonlayer}
      \end{tikzpicture}};
    \end{tikzpicture}
  }
  \caption{\footnotesize \label{fig:closure_2} Stylized example illustrating incoming balance. \textit{Left}: history of two past events $e_1=(t_1,C,\{A,B\})$ and $e_2=(t_2,C,\{D,E\})$. \textit{Right}: a candidate hyperedge $h=(A,\{D,E\})$ for a future hyperevent. The past events $e_1,e_2$ increase the value of incoming balance on $h$ at $t>t_1,t_2$, since the sender of $h$, $A$ received a past event from $C$ and two receivers of $h$, $D$ and $E$ also received a past event from the same sender $C$.} 
\end{figure}

Incoming balance is illustrated in Fig.~\ref{fig:closure_2}. In the notation of that figure, if we ignore the decay over time and if $e_1,e_2$ is the entire history, for a point in time $t>t_1,t_2$ we obtain the following value:
\begin{eqnarray*}
in\_balance_t(h)&=&2/2\enspace.
\end{eqnarray*}

\begin{figure}
  \centering
  \makebox{%
    \begin{tikzpicture}[scale=1,>=stealth]
      \node at (-4,0) (left) {%
      \begin{tikzpicture}[scale=1]
        \tikzstyle{actor}=[circle,minimum size=1mm];
        \node[actor] at (0:1.5) (a) {$A$};
        \node[actor] at (60:1.5) (b) {$B$};
        \node[actor] at (120:1.5) (c) {$C$};
        \node[actor] at (180:1.5) (d) {$D$};
        \node[actor] at (240:1.5) (e) {$E$};
        \node[actor] at (300:1.5) (f) {$F$};
        
        \draw[color=darkgrey,line width=1pt,dashed,->] (a) to (b);
        \draw[color=darkgrey,line width=1pt,dashed,->] (a) to (c);
        \draw[color=darkgrey,line width=1pt,dashed,->] (e) to (d);
        \draw[color=darkgrey,line width=1pt,dashed,->] (e) to (c);

        \begin{pgfonlayer}{background}
          \foreach \nodename in {a,b,c,d,e,f} {
            \coordinate (\nodename') at (\nodename);
          }
          \path[fill=grey1,draw=grey1,line width=0.9cm, line cap=round, line join=round] 
          (a') to (b') to (c') to (a') -- cycle;
          \path[fill=grey3,draw=grey3,line width=0.7cm, line cap=round, line join=round] 
          (c') to (d') to (e') to (c') -- cycle;
        \end{pgfonlayer}
      \end{tikzpicture}};
      \node at (3,0) (right) {%
      \begin{tikzpicture}[scale=1]
        \tikzstyle{actor}=[circle,minimum size=1mm];
        \node[actor] at (0:1.5) (a) {$A$};
        \node[actor] at (60:1.5) (b) {$B$};
        \node[actor] at (120:1.5) (c) {$C$};
        \node[actor] at (180:1.5) (d) {$D$};
        \node[actor] at (240:1.5) (e) {$E$};
        \node[actor] at (300:1.5) (f) {$F$};
        
        \draw[color=darkgrey,line width=1pt,dashed,->] (a) to (d);
        \draw[color=darkgrey,line width=1pt,dashed,->] (a) to (e);

        \begin{pgfonlayer}{background}
          \foreach \nodename in {a,b,c,d,e,f} {
            \coordinate (\nodename') at (\nodename);
          }
          \path[fill=black,draw=black,line width=0.9cm, line cap=round, line join=round] 
          (a') to (d') to (e') to (a') -- cycle;
          \path[fill=white,draw=white,line width=0.83cm, line cap=round, line join=round] 
          (a') to (d') to (e') to (a') -- cycle;
        \end{pgfonlayer}
      \end{tikzpicture}};
    \end{tikzpicture}
  }
  \caption{\footnotesize \label{fig:closure_3} Stylized example illustrating the $shared.receiver$ covariate. \textit{Left}: history of two past events $e_1=(t_1,A,\{B,C\})$ and $e_2=(t_2,E,\{D,C\})$. \textit{Right}: a candidate hyperedge $h=(A,\{D,E\})$ for a future hyperevent. The past events $e_1,e_2$ increase the value of outgoing balance at $t>t_1,t_2$, since the sender of $h$, $A$ has sent a past event to $C$ and one receiver of $h$, $E$ has also sent a past event to $C$.} 
\end{figure}
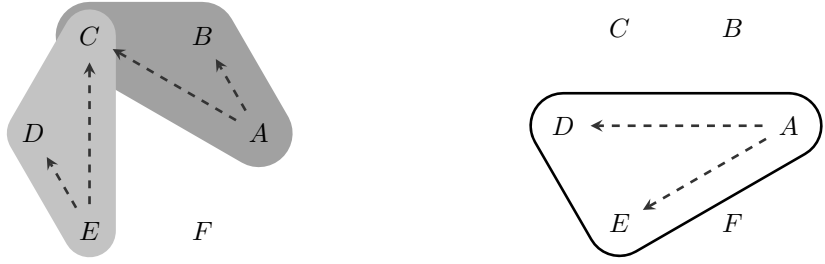

Outgoing balance is illustrated in Fig.~\ref{fig:closure_3}. In the notation of that figure, if we ignore the decay over time and if $e_1,e_2$ is the entire history, for a point in time $t>t_1,t_2$ we obtain the following value:
\begin{eqnarray*}
out\_balance_t(h)&=&1/2\enspace.
\end{eqnarray*}

\subsubsection{Dyadic covariates vs.\ hyperedge covariates}
\label{sec:covar_dyadic_rhem}

As mentioned, we denote a hyperedge covariate $x_t(i,J)$ as a \emph{dyadic covariate} if it admits a decomposition of the form $x_t(i,J)=C(|J|)\cdot\sum_{j\in J}x_t(i,j)$, compare (\ref{eq:covar_dyadic}), where $C(|J|)$ is a multiplicative constant depending only on the size of the receiver set $J$. Here we discuss, which of the hyperedge covariates defined above are dyadic covariates.

Among the covariates based on actor-level attributes, the receiver set average and sender-receiver heterophily are dyadic covariates. In contrast, receiver set heterophily is not a dyadic covariate since the terms in its definition jointly consider pairs of receivers.

Neither exact repetition nor unordered repetition are dyadic covariates in general since the terms in their definition jointly consider a potentially unbounded number of receivers. Partial receiver set repetition of order $p$ is a dyadic covariate for $p=1$ but not for any $p>1$. The same holds for sender-specific partial receiver set repetition of order $p$. Interaction among receivers of order $p$ is not a dyadic covariate for any $p\geq 1$. Note that the definition of this covariate considers subsets of the receiver set $J$ containing $p$ previous receivers and one previous sender. Reciprocation, out-in popularity, and all four triadic closure covariates are dyadic. For the latter, note that in the definition of these covariates we sum over single receivers $j\in J$, rather than over larger subsets of $J$.

\section{Empirical analysis}
\label{sec:experiments}

\subsection{Empirical setting and data}
\label{sec:empirical_data}

We demonstrate the empirical value of the model in an analysis of the Enron email data -- a collection of corporate emails exchanged by employees of Enron Corporation that was made public after the company filed for bankruptcy in December 2001. For additional information on the history of the data, we refer interested readers to \citet{zhou2007strategies}. Prior empirical analyses of the email corpus may be found in \citet{diesner2005communication} and in \citet{keila2005structure}. To facilitate comparability, we analyze the subset of the data cleaned and processed by \citet{zhou2007strategies} that has also been used in the empirical example reported in \citet{perry2013point} and that is available at \url{https://github.com/patperry/interaction-proc/tree/master/data/enron}. Analysis with the \texttt{eventnet} software requires conversion of these data into a different format. The conversion steps are explained in \url{https://github.com/juergenlerner/eventnet/tree/master/data/enron} where the converted data are also available for download. As mentioned before, the entire analysis reported in this paper is detailed in a step-by-step tutorial; see Section~\ref{sec:reproducibility}.

In the empirical section that follows we refer to this subset as the ``Enron data'' or just the ``data.'' The data comprises 21,635 emails (treated as hyperevents) among 156 Enron employees. Additionally, we use the actor attributes, \emph{gender} (female = 1, male = 0), \emph{seniority} (senior = 1, junior = 0), and the categorical attribute \emph{department} (taking the values ``Legal'', ``Trading'', and ``Other''). None of the actor attributes changes over time and there are no missing values.

The observed emails have exactly one sender and between one and 57 receivers. About 30.7\% of the emails have more than one receiver and the average number of receivers is 1.77. The receiver set size distribution is further detailed in Table~\ref{tab:num_receivers}. We analyze all of these emails, that is, we do not discard emails with many receivers.

\begin{table}
  \caption{\footnotesize \label{tab:num_receivers}Number of emails (bottom row) with given number of receivers (top row).}
  \begin{center}
    \begin{tabular}{lrrrrrrrrrrr}
      \hline
      $|J|$ &  1  &   2  &   3  &   4   &  5 &    6  &   7 &    8  &   9  &  10 & $>10$\\
      frequency: &   14,985 &  2,962 & 1,435 &  873 &  711 &  180 &  176  &  61  &  24  &  29 & 199\\
      \hline
  \end{tabular}
  \end{center}
\end{table}

The time values of the emails in the data correspond to seconds -- but time resolution is by the minute since all given time values are divisible by 60. There are 20,994 emails (about 97\%) that have a unique time stamp, there are 305 time points at which exactly two different emails have been sent, nine time points are shared by three emails, and there is one time point at which four different emails have been sent. We order simultaneous emails arbitrarily. Given the high level of time resolution and the relatively low number of simultaneous hyperevents (i.\,e., tied event times), we believe that this decision is unlikely to affect our results in any meaningful way. However, we note that established ways to deal with tied event times exist \citep{kalbfleisch1973marginal,breslow1974covariance,efron1977efficiency,hertz1997validity} and are, for instance, implemented in the \texttt{R} package \texttt{survival}.

\subsection{Model specification and selection}
\label{sec:specification}

The core objectives of our analysis are to understand and illustrate higher order effects in polyadic interaction networks. We  estimate two types of models, a conventional REM using only dyadic covariates (``dyadic model'') and RHEM (``polyadic model'').

The dyadic model includes the covariates receiver set average for the actor-level attributes \emph{gender}, \emph{seniority}, \emph{trading}, and \emph{legal} and sender-receiver heterophily for \emph{gender}, \emph{seniority}, and for the categorical attribute \emph{department}. The dyadic model also includes the network effects partial receiver set repetition and sender-specific partial receiver set repetition of order $p=1$, reciprocation, out-in popularity, and all four triadic closure effects.

The RHEM includes all covariates of the dyadic model and in addition the covariates receiver set heterophily for \emph{gender}, \emph{seniority}, and for the categorical attribute \emph{department}. Moreover, the RHEM includes exact repetition, unordered repetition, as well as (sender-specific) partial receiver set repetition and interaction among receivers for varying values of $p$ that have been determined in a preliminary analysis (see below). Additionally, we specify and estimate two further variants of the dyadic REM and two of the polyadic RHEM. These restricted models include covariates based on actor-level attributes only, and network effects only, respectively.

Model estimation proceeds in two steps. In the first step we apply the \texttt{eventnet} software \citep{lerner2020reliability,lerner2021dynamic}, available at \url{https://github.com/juergenlerner/eventnet}, to sample $k$ non-event hyperedges associated with each observed hyperevent and to compute (a superset of) all covariates of observed events and sampled non-events. We set the number of non-events per event to $k=100$. We initially compute (sender-specific) partial receiver set repetition and interaction among receivers of order $p=1,\dots,10$. All network effects are defined with the half life period $T_{1/2}$ set to one week. (While the decision about the half life is essentially arbitrary, and unrelated to the objectives of this paper, it can be expected that one week somewhat levels out day-of-the-week effects, which are likely to be prominent in a corporate communication network.) In the second step we estimate parameters of models specified by varying lists of covariates with the \texttt{coxph} function in the \texttt{R} package \texttt{survival} \citep{therneau2013modeling,therneau2015survival}. To assess the variation of estimates over different samples, we repeat the sampling of risk sets 100 times, recomputing the covariate values each time. This gives us, for each actual choice of the covariate vector $x_t$, 100 different log partial likelihood functions (\ref{eq:likelihood_rhem_sampled}) and 100 potentially different estimated parameter vectors $\hat{\beta}$. Since covariates modeling network effects are skewed, we transform them by $x\mapsto\sqrt{x}$, but we do not standardize covariates.

To determine reasonable values of $p$ for the covariates (sender-specific) partial receiver set repetition and interaction among receivers, we incrementally add these covariates for increasing values of $p$ and monitor robustness of parameter estimation over different samples, as well as model fit assessed by information criteria (AIC and BIC). 
We find that for a model specified with partial receiver set repetition up to order $p=4$ and sender-specific partial receiver set repetition and interaction among receivers up to order $p=3$, estimation converges for all 100 samples and model fit measured in AIC and BIC consistently increases for growing order of these effects. In the following we denote this model with 28 covariates as ``the RHEM'' (see the rightmost model in Table~\ref{tab:models_rhem}).

Among the covariates included in the RHEM, two pairs have correlation exceeding $0.9$. These are the pairs exact repetition and unordered repetition, with a correlation of $0.93$, and partial receiver set repetition and sender-specific partial receiver set repetition of order $p=3$, with a correlation equal to $0.98$. To test how findings are affected by such high correlations, we additionally fit a reduced RHEM without unordered repetition and without sender-specific partial receiver set repetition of order $p=3$. This reduced RHEM has a lower model fit than the RHEM including all effects ($AIC=73,286$ for the reduced model and $AIC=72,919$ for the largest RHEM). Most parameters that are included in both models are not affected qualitatively in the sense that they keep their signs and significance levels. The only exceptions are (1) out-in popularity becomes significantly positive in the reduced model (and is non-significant in the largest RHEM) and (2) the parameter of exact repetition switches its sign from significantly positive to significantly negative when we additionally include unordered repetition. We argue that this sign switch does not point to lack of robustness in our findings -- but rather to a relevant effect in multicast communication networks that can be well explained (see the discussion related with exact repetition and unordered repetition below).

\section{Results and discussion}
\label{sec:results}

\subsection{Discussion of effects}
\label{sec:results_effects}

Estimated parameters and standard errors of the three models specified with dyadic covariates are reported in Table~\ref{tab:models_dyadic}, those of RHEM including dyadic and higher-order effects are reported in Table~\ref{tab:models_rhem}. In the following we discuss findings to highlight representative effects, or types of effects, across all models in which the respective covariates are included. Our goal is not to discuss the findings exhaustively, but rather to offer guidance how to interpret effects and illustrate the additional results that RHEM make possible, and the new questions they afford. Even though we fitted models to 100 different samples, the reported parameters have been estimated from one arbitrarily selected single sample. Indeed, we believe that this situation is more representative for empirical studies than results derived from repeated sampling. We report summary statistics of the distribution of parameters over 100 samples in Appendix~\ref{app:variation}. We emphasize that dyadic REM and RHEM are estimated on the same sample; the only difference is in the included effects.

\begin{table}
\caption{\footnotesize \label{tab:models_dyadic}Estimated parameters of dyadic models.}
\begin{center}
    \begin{tabular}{l r r r }
\hline
 & att (dyad) & net (dyad) & att+net (dyad) \\
\hline
rec\_avg\_female               & $0.24 \; (0.02)^{***}$  &                         & $0.26 \; (0.02)^{***}$  \\
rec\_avg\_senior            & $0.71 \; (0.02)^{***}$  &                         & $0.45 \; (0.02)^{***}$  \\
rec\_avg\_legal             & $1.22 \; (0.02)^{***}$  &                         & $0.15 \; (0.03)^{***}$  \\
rec\_avg\_trading           & $-0.61 \; (0.03)^{***}$ &                         & $-0.22 \; (0.03)^{***}$ \\
send\_rec\_diff\_female      & $-0.54 \; (0.02)^{***}$ &                         & $-0.24 \; (0.02)^{***}$ \\
send\_rec\_diff\_senior   & $-1.04 \; (0.02)^{***}$ &                         & $-0.49 \; (0.02)^{***}$ \\
send\_rec\_diff\_dept & $-2.10 \; (0.02)^{***}$ &                         & $-0.88 \; (0.02)^{***}$ \\
rec\_sub\_rep\_1                  &                         & $0.11 \; (0.01)^{***}$  & $0.04 \; (0.01)^{**\phantom{*}}$   \\
send\_rec\_sub\_rep\_1                  &                         & $2.70 \; (0.02)^{***}$  & $2.54 \; (0.02)^{***}$  \\
reciprocation                  &                         & $1.04 \; (0.02)^{***}$  & $1.00 \; (0.02)^{***}$  \\
out\_in\_pop                  &                         & $-0.00 \; (0.01)\phantom{^{***}}$       & $0.01 \; (0.01)\phantom{^{***}}$        \\
in\_balance                &                         & $0.11 \; (0.01)^{***}$  & $0.10 \; (0.01)^{***}$  \\
out\_balance              &                         & $-0.02 \; (0.01)^{*\phantom{**}}$   & $-0.04 \; (0.01)^{***}$ \\
transitive\_closure           &                         & $0.12 \; (0.01)^{***}$  & $0.11 \; (0.01)^{***}$  \\
cyclic\_closure               &                         & $-0.09 \; (0.01)^{***}$ & $-0.11 \; (0.01)^{***}$ \\
\hline
AIC                          & $164871.80$             & $87983.89$              & $85158.09$              \\
Num. events                  & 21,635                   & 21,635                   & 21,635                   \\
Num. obs.                    & 2,185,135                 & 2,185,135                 & 2,185,135                 \\
\hline
\multicolumn{4}{l}{\scriptsize{$^{***}p<0.001$, $^{**}p<0.01$, $^*p<0.05$}}
    \end{tabular}
    \end{center}
\end{table}

\begin{table}
\caption{\footnotesize \label{tab:models_rhem}Estimated parameters of RHEM. }
    \begin{center}
\begin{tabular}{l r r r }
\hline
 & att (rhem) & net (rhem) & att+net (rhem) \\
\hline
rec\_avg\_female                & $0.28 \; (0.02)^{***}$  &                         & $0.21 \; (0.02)^{***}$  \\
rec\_avg\_senior             & $0.68 \; (0.02)^{***}$  &                         & $0.32 \; (0.02)^{***}$  \\
rec\_avg\_legal              & $1.23 \; (0.02)^{***}$  &                         & $0.13 \; (0.03)^{***}$  \\
rec\_avg\_trading            & $-0.49 \; (0.03)^{***}$ &                         & $-0.11 \; (0.03)^{***}$ \\
send\_rec\_diff\_female       & $-0.50 \; (0.02)^{***}$ &                         & $-0.19 \; (0.02)^{***}$ \\
send\_rec\_diff\_senior    & $-0.94 \; (0.02)^{***}$ &                         & $-0.42 \; (0.02)^{***}$ \\
send\_rec\_diff\_dept  & $-1.88 \; (0.02)^{***}$ &                         & $-0.73 \; (0.02)^{***}$ \\
rec\_set\_diff\_female      & $-0.43 \; (0.04)^{***}$ &                         & $-0.18 \; (0.07)^{**\phantom{*}}$  \\
rec\_set\_diff\_senior   & $-1.42 \; (0.04)^{***}$ &                         & $-0.65 \; (0.07)^{***}$ \\
rec\_set\_diff\_dept & $-2.22 \; (0.04)^{***}$ &                         & $-0.99 \; (0.07)^{***}$ \\
exact\_repetition                    &                         & $-0.49 \; (0.06)^{***}$ & $-0.37 \; (0.06)^{***}$ \\
unordered\_repetition         &                         & $1.04 \; (0.05)^{***}$  & $0.98 \; (0.05)^{***}$  \\
rec\_sub\_rep\_1                   &                         & $0.07 \; (0.01)^{***}$  & $0.02 \; (0.01)\phantom{^{***}}$        \\
rec\_sub\_rep\_2                   &                         & $0.65 \; (0.06)^{***}$  & $0.47 \; (0.06)^{***}$  \\
rec\_sub\_rep\_3                   &                         & $2.18 \; (0.27)^{***}$  & $1.93 \; (0.26)^{***}$  \\
rec\_sub\_rep\_4                   &                         & $6.73 \; (0.80)^{***}$  & $6.12 \; (0.79)^{***}$  \\
send\_rec\_sub\_rep\_1                   &                         & $1.99 \; (0.04)^{***}$  & $1.80 \; (0.04)^{***}$  \\
send\_rec\_sub\_rep\_2                   &                         & $5.08 \; (0.21)^{***}$  & $4.83 \; (0.21)^{***}$  \\
send\_rec\_sub\_rep\_3                   &                         & $5.15 \; (1.01)^{***}$  & $4.03 \; (0.92)^{***}$  \\
reciprocation                   &                         & $0.65 \; (0.03)^{***}$  & $0.63 \; (0.03)^{***}$  \\
out\_in\_pop                   &                         & $-0.01 \; (0.01)\phantom{^{***}}$       & $0.01 \; (0.01)\phantom{^{***}}$        \\
interact\_rec\_1            &                         & $2.89 \; (0.09)^{***}$  & $2.67 \; (0.09)^{***}$  \\
interact\_rec\_2            &                         & $7.30 \; (0.88)^{***}$  & $6.93 \; (0.87)^{***}$  \\
interact\_rec\_3            &                         & $34.90 \; (5.15)^{***}$ & $32.74 \; (5.21)^{***}$ \\
in\_balance                 &                         & $0.08 \; (0.01)^{***}$  & $0.08 \; (0.01)^{***}$  \\
out\_balance               &                         & $-0.09 \; (0.01)^{***}$ & $-0.10 \; (0.01)^{***}$ \\
transitive\_closure            &                         & $0.06 \; (0.01)^{***}$  & $0.05 \; (0.01)^{***}$  \\
cyclic\_closure                &                         & $-0.04 \; (0.01)^{**\phantom{*}}$  & $-0.05 \; (0.01)^{***}$ \\
\hline
AIC                           & $159300.98$             & $75196.57$              & $72919.66$              \\
Num. events                   & 21,635                   & 21,635                   & 21,635                   \\
Num. obs.                     & 2,185,135                 & 2,185,135                 & 2,185,135                 \\
\hline
\multicolumn{4}{l}{\scriptsize{$^{***}p<0.001$, $^{**}p<0.01$, $^*p<0.05$}}
\end{tabular}
    \end{center}
\end{table}

\paragraph{Effects of actor-level attributes.} The attribute distribution in receiver sets is captured by the four covariates measuring receiver set averages ($rec\_avg\_z$). All four models including attribute effects agree that receiver sets have an over-representation of females, seniors, and members of the legal department and an under-representation of members of the trading department (recall that the base, or reference, department is ``other''). Parameters of covariates assessing the difference in attribute values between the sender and the receivers consistently reveal that senders tend to write messages to receivers of the same gender, seniority and department. On top of this ``sender-receiver homophily'', the two RHEM containing attribute effects reveal that receiver sets tend to be homogeneous with respect to gender, seniority, and department (negative parameters of the ``receiver set difference'' covariates). We observe that without any exception effect sizes of covariates based on attributes become smaller when we additionally control for network effects. Moreover, typically -- but not without exception -- sizes of attribute effects are stronger in the dyadic models than in the RHEM. Intuitively, these results reveal a possible tendency to over-estimate the effects of actor-specific attributes in models that do not control for network effects, or that do not control for higher-order effects, respectively.

\paragraph{Exact repetition and unordered repetition.} In the RHEM the parameter associated with unordered repetition is significantly positive and that of exact repetition is significantly negative. These two effects have to be interpreted together -- as we illustrate building on the example from Fig.~\ref{fig:repetition}. Assume that at $t_1$ actor $A$ sends an email to $\{B,C,D,E\}$ and consider two alternative hyperedges $h=(C,\{A,B,D,E\})$ and $h'=(A,\{B,C,D,E\})$ at a later point in time $t>t_1$. The past email $e_1=(t_1,A,\{B,C,D,E\})$ increases the value of exact repetition and it increases the value of unordered repetition on $h'$ at $t$. The joint effect of these two covariates is positive (since unordered repetition has a larger parameter than the absolute value of the negative repetition parameter). Thus, the past event $e_1=(t_1,A,\{B,C,D,E\})$ makes a repeated email from $A$ to $\{B,C,D,E\}$ at $t>t_1$ more likely. However, an event at $t$ on hyperedge $h$, that is, sent from $C$ to $\{A,B,D,E\}$ would be even more likely than an email from $A$ to $\{B,C,D,E\}$. This is because the past email $e_1=(t_1,A,\{B,C,D,E\})$ increases the value of unordered repetition on $h$ at $t$ (having a positive effect) but it does not increase the value of repetition (which would have had a negative effect).

Thus, while conversations within fixed lists of actors (e.\,g., $A,B,C,D,E$ in the example from Fig.~\ref{fig:repetition}) are overrepresented, it is more likely that a future email within the same fixed set of actors will have a sender different from that of the preceding email. This points to a form of ``turn-taking'' \citep{gibson2005taking} within fixed conversation groups. We note that it would not be possible to express these effects purely with dyadic covariates.

Exact repetition and unordered repetition express a form of behavioral inertia, indicating that actors tend to repeat what they did in the past.

\paragraph{(Sender-specific) partial receiver set repetition and interaction among receivers.} In the RHEM almost all covariates expressing subset repetition in the receiver set (with order $p=1,2,3,4$), all covariates for sender-specific subset repetition in the receiver set (with order $p=1,2,3$), and interaction among receivers have a significantly positive effect. (Sender-specific) receiver subset repetition of order one are dyadic effects and thus are also included in the dyadic model, where they are associated with positive parameters.

All three types of effects can be interpreted in defining a similarity (or suitability) measure on sets of actors based on past interaction, which in turn tends to increase the rate of future interactions send by actor $i$ to receiver set $J$. Specifically, partial receiver set repetition considers $J$ as a suitable receiver set to the extent that subsets $J'\subseteq J$ have already received common interactions; sender-specific partial receiver set repetition considers the hyperedge $(i,J)$ as suitable for the next interaction to the extent that subsets $J'\subseteq J$ have already received common interactions from the sender $i$; the interaction among receivers effect considers $J$ as a suitable receiver set to the extent that subsets $J'\subseteq J$ have already received common interactions from a sender $i'\in J\setminus J'$. In all three types of effects, the order of the effect corresponds to the size of the subsets $J'$.  

The findings demonstrate that repetition from the same sender to the same receivers (possibly within a larger and varying set of yet other receivers) is not a purely dyadic phenomenon. Instead, if two (or three or four) actors have already jointly received an email, then they are more likely to do so again. This points to a clustering of the set of all possible receivers into subsets that are likely to receive the same emails. We find evidence for a ``global'' clustering (that is, one that applies to the average sender) and for a sender-specific clustering (so that one sender can structure the space of receivers in a different way than another sender).

\paragraph{Reciprocation and out-in popularity.} A tendency to reciprocate interaction is found in all four models containing this effect. Thus, receivers of past interaction have the tendency to send interaction to the sender of that past interaction. Beyond that effect, all four models agree that there is no evidence for out-in popularity. This latter effect tests whether those actors who have sent out more interaction in the past are likely to receive more future interaction -- from anyone in the network, that is, not necessarily from the receivers of the past interaction.

\paragraph{Triadic effects.} The triadic effects can be included in the dyadic model and in the RHEM. All four models agree that there is a tendency for transitive closure but a tendency against cyclic closure. Together these two findings are compatible with the ``hierarchical'' interpretation that messages tend to go from higher status to lower status actors \citep{lerner2017third}. Note that these findings are equally compatible with the interpretation that messages tend to go ``upward'' in the hierarchy. We also find that two actors who have received (possibly different) messages from the same third actor (``incoming balance'') are more likely to interact themselves but two actors who have both sent messages to the same third actor (``outgoing balance'') are less likely to interact themselves.

\subsection{Contribution of individual covariates to the log-likelihood}
\label{sec:covariate_contributions}

In Table~\ref{tab:cov_contribs_to_null} we list improvements in log likelihood of the 28 models specified with exactly one covariate over the null model (i.\,e., the model with no covariates). We also report the contribution of individual covariates in the full model by measuring improvements of the full model (i.\,e., the ``RHEM'' with 28 covariates) over the 28 models obtained by dropping exactly one covariate. These latter measurements indicate how much individual covariates contribute over the 27 others.

\begin{table}
\caption{\footnotesize \label{tab:cov_contribs_to_null}Improvement in log-likelihood resulting from individual covariates. }
    \begin{center}
\begin{tabular}{lrr}
  \hline
 & over null model & in full model \\ 
  \hline
send\_rec\_sub\_rep\_1 & 53649.62 & 1247.12 \\ 
  unordered\_repetition & 47906.19 & 173.16 \\ 
  exact\_repetition & 44603.05 & 19.00 \\ 
  reciprocation & 36047.57 & 169.54 \\ 
  send\_rec\_sub\_rep\_2 & 22539.50 & 530.31 \\ 
  transitive\_closure & 20468.45 & 10.09 \\ 
  out\_balance & 18562.00 & 57.88 \\ 
  rec\_sub\_rep\_2 & 18444.34 & 25.76 \\ 
  cyclic\_closure & 17300.53 & 6.30 \\ 
  in\_balance & 16599.53 & 43.17 \\ 
  rec\_sub\_rep\_1 & 16274.49 & 1.46 \\ 
  interact\_rec\_1 & 15086.75 & 433.19 \\ 
  rec\_sub\_rep\_3 & 13251.15 & 24.83 \\ 
  send\_rec\_sub\_rep\_3 & 13089.58 & 13.23 \\ 
  send\_rec\_diff\_dept & 11048.89 & 524.04 \\ 
  interact\_rec\_2 & 10794.70 & 46.04 \\ 
  out\_in\_pop & 10294.25 & 0.07 \\ 
  rec\_sub\_rep\_4 & 8382.27 & 34.92 \\ 
  interact\_rec\_3 & 6614.91 & 28.90 \\ 
  rec\_avg\_legal & 5599.23 & 7.55 \\ 
  rec\_set\_diff\_dept & 2430.60 & 103.30 \\ 
  send\_rec\_diff\_senior & 2250.83 & 183.04 \\ 
  rec\_avg\_trading & 1955.37 & 8.20 \\ 
  rec\_set\_diff\_senior & 1071.38 & 43.45 \\ 
  rec\_avg\_senior & 1001.24 & 92.09 \\ 
  send\_rec\_diff\_female & 779.91 & 33.37 \\ 
  rec\_avg\_female & 461.45 & 38.04 \\ 
  rec\_set\_diff\_female & 70.88 & 3.37 \\ 
   \hline
\end{tabular}
    \end{center}
\end{table}

We find that sender-specific partial receiver set repetition of order one (which is a dyadic covariate) is the strongest individual effect, closely followed by unordered repetition and exact repetition (both of which are non-dyadic hyperedge covariates). The covariates based on actor-level attributes imply much lower contributions to the log likelihood than the network effects. Looking at the contributions of single effects on top of all other 27 covariates (rightmost column) we find -- with exception of the strongest effect -- a remarkably different order. However, the second strongest effect, sender-specific partial receiver set repetition of order two, is again a non-dyadic hyperedge effect. We conclude that the contributions of some of the higher-order effects are roughly comparable to those of the strongest dyadic effects.

\subsection{Evidence of higher-order effects}
\label{sec:higher_order}

Overall, we find strong evidence of the presence of higher-order effects in the data. Covariates that go beyond purely dyadic effects have been found to be significantly predictive of future event distributions. Moreover, as shown above, the contributions of some higher-order effects to the log likelihood have similar magnitudes as the strongest dyadic covariates.

As outlined above, we contend that such interdependence between different receivers should not be considered as an annoyance to be controlled away -- but rather as an opportunity to develop and test additional theories about the structure of polyadic interaction networks. Some of the higher-order effects are highly relevant for empirical studies. Homophily in the receiver set, exact repetition and unordered repetition, and (sender-specific) partial receiver-set repetition of order $p\geq 2$ all have relevant implications for the structure and dynamics of polyadic interaction networks. 

\subsection{Qualitative differences between RHEM and dyadic models}
\label{sec:differences}

In most cases, parameters of covariates that have been included in dyadic models and in RHEM are estimated to have the same sign and significance levels. However, in most cases the absolute values of these parameters are considerably smaller in the RHEM than in the dyadic models. This may suggest that failure to control for higher-order effects could lead to an overestimation of effect sizes. In our empirical analysis the dyadic models suggested that partial receiver set repetition of order one is significantly positive, while the largest RHEM found the same effect to be not significant. 

\subsection{Model fit of RHEM and dyadic model}
\label{sec:model_fit}

Regarding model fit, measured with information criteria (AIC), we find two patterns. First, network effects improve the model fit much more than attribute effects and the joint models including attribute and network effects have the best fit in the family of dyadic models and in RHEMs. Second, RHEM fit the data better than the corresponding dyadic models. We find that the RHEM purely specified with network effects ``net (rhem)'' already has a better fit than the full dyadic model ``att+net (dyad)''.

\subsection{Replicability}
\label{sec:reproducibility}

The analysis is explained in a tutorial linked from \url{https://github.com/juergenlerner/eventnet/wiki/}, from the data preprocessing and the computation of covariates over to model estimation in R. Thus, the analysis reported in this paper is fully replicable and the tutorial points out some model variations that might be relevant in given application scenarios. The software  may be adopted in -- and adapted to future empirical studies involving polyadic social interaction processes.

\section{Conclusion}
\label{sec:conclusion}

The objectives of our paper are to introduce, analyze, and illustrate hyperedge covariates for polyadic interaction networks and to compare RHEM, including higher-order effects, with REM purely specified in terms of dyadic covariates. In our illustrative empirical analysis of the Enron email data, we find consistent evidence for the significance and importance of higher-order effects captured by hyperedge covariates that jointly consider receiver sets of size larger than one. Not controlling for such higher-order dependencies can also affect estimated parameters of dyadic covariates, although, in our analysis, this change seems to affect mostly parameter sizes (and occasionally also their significance levels), but in most cases does not switch the direction of effects (i.\,e., parameter signs). 

The results we reported suggest that researchers interested in analyzing polyadic interaction networks should not restrict their models to dyadic specifications. From a high-level perspective, the step from models for multicast networks specified via dyadic covariates \citep{perry2013point} to RHEM is relatively intuitive. However, hyperedge covariates provide a much richer set of possible effects and considerably increase the fit of models for the given empirical data.

These results strengthen our view that that higher-order effects should not be considered merely as an annoyance that has to be controlled for -- but rather as an opportunity to develop and test additional theories about the structure and dynamics of social interaction and communication networks. Some of the effects that we have documented could not have been discovered or tested with available models purely specified via dyadic covariates. Concretely, this includes findings on unordered repetition (pointing to turn-taking within a fixed list of conversation participants), partial receiver set repetition (related with a clustering of the actors into groups that are likely receivers of a joint message), and sender-specific partial receiver set repetition (pointing to patterns in which different senders cluster the actors into potentially different groups). 
 
We point out that average hyperedge sizes in the Enron email data are still rather small: 70\% of the emails have exactly one receiver and the average number of receivers is 1.77. In other empirical settings, for instance, coauthorship networks \citep{newman2004coauthorship}, hyperevents are typically much larger. For example, in the coauthorship networks considered in \citet{lerner2019rem}, the average number of authors per paper is close to eight and some events have size up to 100. As another example, in the meeting events extracted from contact diaries analyzed in \citet{lerner2021dynamic}, the average number of participants per meeting is 4.4. With larger hyperevents, it is possible to include subset repetition covariates of higher order.

Our primary goal in this paper was to establish hyperedge covariates and compare RHEM to dyadic models, rather than elaborate new parameter estimation techniques. To deal with the exponential size of the risk set, we proposed case-control sampling -- which is a readily available and well-established technique in the analysis of rare events \citep{bgl-mascdcphm-95,keogh2014nested}. Hyperevents could indeed be considered as rare events since a randomly selected subset of actors is very unlikely to experience even one common event. However, case-control sampling, in which we uniformly sample from the risk set, revealed some limitations, mostly due to the fact that some higher-order covariates are very sparse in the risk set. In turn, most sampled non-events did not provide any information on those covariates which increased the necessary sample size. A more efficient sampling scheme could be stratified sampling, where, for instance, one could include as strata non-events that are non-zero in some subset repetition covariates. Another possibility would be Markov-chain Monte Carlo (MCMC) methods which do not sample uniformly from the (stratified or unstratified) risk set but sample proportional to the conditional event probability of subsets. The additional cost of MCMC methods, however, is that sampling is no longer independent of model parameters.

We point out that, even if we do not include subset repetition covariates of order five or higher, our models do not preclude hyperevents of that size. The reason for this is that subset repetition covariates are nested within each other: whenever a past event implies a non-zero value in subset repetition of order $p$, then it necessarily implies non-zero values in subset repetition of lower order $p'<p$; compare the examples given in Section~\ref{sec:covar_network}.

Besides improved parameter estimation techniques for RHEM, we consider the further development of other RHEM covariates as a promising direction for future work. As we have discussed,  the collection of hyperedge covariates proposed in this paper is far from exhausting the possibilities. Developing further effects that are relevant for given application scenarios will further increase the empirical extension and generality of RHEM.

\section*{Acknowledgements}
This work has been supported by the Deutsche Forschungsgemeinschaft (DFG) -- grant no.\ 321869138.


\begin{thebibliography}{45}
\newcommand{\enquote}[1]{``#1''}
\expandafter\ifx\csname natexlab\endcsname\relax\def\natexlab#1{#1}\fi

\bibitem[{Ahmadpoor and Jones(2019)}]{ahmadpoor2019decoding}
Ahmadpoor, M. and Jones, B.~F. (2019), \enquote{Decoding team and individual
  impact in science and invention,} \textit{Proceedings of the National Academy
  of Sciences}, 201812341.

\bibitem[{Amati et~al.(2019)Amati, Lomi, and Mascia}]{amati2019some}
Amati, V., Lomi, A., and Mascia, D. (2019), \enquote{Some days are better than
  others: Examining time-specific variation in the structuring of
  interorganizational relations,} \textit{Social Networks}, 57, 18--33.

\bibitem[{Berge(1989)}]{berge1989hypergraphs}
Berge, C. (1989), \textit{Hypergraphs: combinatorics of finite sets},
  North-Holland.

\bibitem[{Bianchi and Lomi(2022)}]{bianchi2022fromties}
Bianchi, F. and Lomi, A. (2022), \enquote{From Ties to Events in the Analysis
  of Interorganizational Exchange Relations,} \textit{Organizational Research
  Methods}, Forthcoming.

\bibitem[{Borgan et~al.(1995)Borgan, Goldstein, and
  Langholz}]{bgl-mascdcphm-95}
Borgan, {\O}., Goldstein, L., and Langholz, B. (1995), \enquote{Methods for the
  analysis of sampled cohort data in the {C}ox proportional hazards model,}
  \textit{The Annals of Statistics}, 1749--1778.

\bibitem[{Brandes et~al.(2009)Brandes, Lerner, and
  Snijders}]{brandes2009networks}
Brandes, U., Lerner, J., and Snijders, T.~A. (2009), \enquote{Networks evolving
  step by step: statistical analysis of dyadic event data,} in \textit{Proc.\
  intl.\ conf.\ Advances in Social Network Analysis and Mining (ASONAM)}, IEEE
  Computer Society, pp. 200--205.

\bibitem[{Breslow(1974)}]{breslow1974covariance}
Breslow, N. (1974), \enquote{Covariance analysis of censored survival data,}
  \textit{Biometrics}, 30, 89--99.

\bibitem[{Bronsteen and Fiss(2002)}]{bronsteen2002class}
Bronsteen, J. and Fiss, O. (2002), \enquote{The class action rule,}
  \textit{Notre Dame L. Rev.}, 78, 1419.

\bibitem[{Butts(2008)}]{butts2008relational}
Butts, C.~T. (2008), \enquote{A relational event framework for social action,}
  \textit{Sociological Methodology}, 38, 155--200.

\bibitem[{Colizza et~al.(2007)Colizza, Barrat, Barthelemy, Valleron, and
  Vespignani}]{colizza2007modeling}
Colizza, V., Barrat, A., Barthelemy, M., Valleron, A.-J., and Vespignani, A.
  (2007), \enquote{Modeling the worldwide spread of pandemic influenza:
  baseline case and containment interventions,} \textit{PLoS medicine}, 4, e13.

\bibitem[{Diesner et~al.(2005)Diesner, Frantz, and
  Carley}]{diesner2005communication}
Diesner, J., Frantz, T.~L., and Carley, K.~M. (2005), \enquote{Communication
  networks from the {Enron} email corpus "It's always about the people. {Enron}
  is no different",} \textit{Computational \& Mathematical Organization
  Theory}, 11, 201--228.

\bibitem[{Efron(1977)}]{efron1977efficiency}
Efron, B. (1977), \enquote{The efficiency of Cox's likelihood function for
  censored data,} \textit{Journal of the American statistical Association}, 72,
  557--565.

\bibitem[{Fowler et~al.(2007)Fowler, Johnson, Spriggs, Jeon, and
  Wahlbeck}]{fowler2007network}
Fowler, J.~H., Johnson, T.~R., Spriggs, J.~F., Jeon, S., and Wahlbeck, P.~J.
  (2007), \enquote{Network analysis and the law: Measuring the legal importance
  of precedents at the US Supreme Court,} \textit{Political Analysis}, 15,
  324--346.

\bibitem[{Freeman(2003)}]{freeman2003finding}
Freeman, L.~C. (2003), \enquote{Finding social groups: A meta-analysis of the
  {S}outhern {W}omen data,} in \textit{Dynamic Social Network Modeling and
  Analysis: Workshop Summary and Papers}, eds. Breiger, R., Carley, C., and
  Pattison, P., Washington, DC: National Research Council, The National
  Academies Press, pp. 39--97.

\bibitem[{Gibson(2005)}]{gibson2005taking}
Gibson, D.~R. (2005), \enquote{Taking turns and talking ties: Networks and
  conversational interaction,} \textit{American journal of sociology}, 110,
  1561--1597.

\bibitem[{Guimera et~al.(2005)Guimera, Uzzi, Spiro, and
  Amaral}]{guimera2005team}
Guimera, R., Uzzi, B., Spiro, J., and Amaral, L. A.~N. (2005), \enquote{Team
  assembly mechanisms determine collaboration network structure and team
  performance,} \textit{Science}, 308, 697--702.

\bibitem[{H{\^a}ncean et~al.(2021)H{\^a}ncean, Lerner, Perc, Ghi{\c{t}}{\u{a}},
  Bunaciu, Stoica, and Mih{\u{a}}il{\u{a}}}]{hancean2021role}
H{\^a}ncean, M.-G., Lerner, J., Perc, M., Ghi{\c{t}}{\u{a}}, M.~C., Bunaciu,
  D.-A., Stoica, A.~A., and Mih{\u{a}}il{\u{a}}, B.-E. (2021), \enquote{The
  role of age in the spreading of {COVID-19} across a social network in
  {B}ucharest,} \textit{Journal of Complex Networks}, 9.

\bibitem[{Hertz-Picciotto and Rockhill(1997)}]{hertz1997validity}
Hertz-Picciotto, I. and Rockhill, B. (1997), \enquote{Validity and efficiency
  of approximation methods for tied survival times in Cox regression,}
  \textit{Biometrics}, 53, 1151--1156.

\bibitem[{Hollway and Koskinen(2016)}]{hollway2016multilevel}
Hollway, J. and Koskinen, J. (2016), \enquote{Multilevel bilateralism and
  multilateralism: States' bilateral and multilateral fisheries treaties and
  their secretariats,} in \textit{Multilevel Network Analysis for the Social
  Sciences}, Springer, pp. 315--332.

\bibitem[{Kalbfleisch and Prentice(1973)}]{kalbfleisch1973marginal}
Kalbfleisch, J.~D. and Prentice, R.~L. (1973), \enquote{Marginal likelihoods
  based on Cox's regression and life model,} \textit{Biometrika}, 60, 267--278.

\bibitem[{Keila and Skillicorn(2005)}]{keila2005structure}
Keila, P.~S. and Skillicorn, D.~B. (2005), \enquote{Structure in the {E}nron
  email dataset,} \textit{Computational \& Mathematical Organization Theory},
  11, 183--199.

\bibitem[{Keogh and Cox(2014)}]{keogh2014nested}
Keogh, R.~H. and Cox, D.~R. (2014), \textit{Nested case-control studies},
  Cambridge University Press, chap.~7, Institute of Mathematical Statistics
  Monographs, pp. 160--190.

\bibitem[{Kim et~al.(2018)Kim, Schein, Desmarais, and
  Wallach}]{kim2018hyperedge}
Kim, B., Schein, A., Desmarais, B.~A., and Wallach, H. (2018), \enquote{The
  Hyperedge Event Model,} \textit{arXiv preprint arXiv:1807.08225}.

\bibitem[{Kuhn et~al.(2020)Kuhn, Younge, and Marco}]{kuhn2020patent}
Kuhn, J., Younge, K., and Marco, A. (2020), \enquote{Patent citations
  reexamined,} \textit{The RAND Journal of Economics}, 51, 109--132.

\bibitem[{Lazer et~al.(2009)Lazer, Pentland, Adamic, Aral, Barabasi, Brewer,
  Christakis, Contractor, Fowler, Gutmann, et~al.}]{lazer2009social}
Lazer, D., Pentland, A., Adamic, L., Aral, S., Barabasi, A.-L., Brewer, D.,
  Christakis, N., Contractor, N., Fowler, J., Gutmann, M., et~al. (2009),
  \enquote{Computational social science.} \textit{Science}, 323, 721--723.

\bibitem[{Leenders et~al.(2016)Leenders, Contractor, and
  DeChurch}]{leenders2016once}
Leenders, R. T.~A., Contractor, N.~S., and DeChurch, L.~A. (2016),
  \enquote{Once upon a time: Understanding team processes as relational event
  networks,} \textit{Organizational Psychology Review}, 6, 92--115.

\bibitem[{Lerner et~al.(2013)Lerner, Bussmann, Snijders, and
  Brandes}]{lerner2013modeling}
Lerner, J., Bussmann, M., Snijders, T.~A., and Brandes, U. (2013),
  \enquote{Modeling frequency and type of interaction in event networks,}
  \textit{Corvinus Journal of Sociology and Social Policy}, 4, 3--32.

\bibitem[{Lerner and Lomi(2017)}]{lerner2017third}
Lerner, J. and Lomi, A. (2017), \enquote{The third man: Hierarchy formation in
  Wikipedia,} \textit{Applied network science}, 2, 1--30.

\bibitem[{Lerner and Lomi(2020)}]{lerner2020reliability}
Lerner, J. and Lomi, A. (2020), \enquote{Reliability of relational event model estimates under
  sampling: How to fit a relational event model to 360 million dyadic events,}
  \textit{Network Science}, 8, 97--135.

\bibitem[{Lerner and Lomi(2022)}]{lerner2022dynamic}
Lerner, J. and Lomi, A. (2022), \enquote{A dynamic model for the mutual constitution of individuals
  and events,} \textit{Journal of Complex Networks}, 10, cnac004.

\bibitem[{Lerner et~al.(2021)Lerner, Lomi, Mowbray, Rollings, and
  Tranmer}]{lerner2021dynamic}
Lerner, J., Lomi, A., Mowbray, J., Rollings, N., and Tranmer, M. (2021),
  \enquote{Dynamic network analysis of contact diaries,} \textit{Social
  Networks}, 66, 224--236.

\bibitem[{Lerner et~al.(2019)Lerner, Tranmer, Mowbray, and
  Hancean}]{lerner2019rem}
Lerner, J., Tranmer, M., Mowbray, J., and Hancean, M.-G. (2019), \enquote{REM
  beyond dyads: relational hyperevent models for multi-actor interaction
  networks,} \textit{arXiv preprint arXiv:1912.07403},
  \url{https://arxiv.org/abs/1912.07403}.

\bibitem[{Mulder and Hoff(2021)}]{mulder2021latent}
Mulder, J. and Hoff, P.~D. (2021), \enquote{A Latent Variable Model for
  Relational Events with Multiple Receivers,} \textit{arXiv preprint
  arXiv:2101.05135}.

\bibitem[{Newman(2004)}]{newman2004coauthorship}
Newman, M.~E. (2004), \enquote{Coauthorship networks and patterns of scientific
  collaboration,} \textit{Proceedings of the national academy of sciences},
  101, 5200--5205.

\bibitem[{Perry and Wolfe(2013)}]{perry2013point}
Perry, P.~O. and Wolfe, P.~J. (2013), \enquote{Point process modelling for
  directed interaction networks,} \textit{Journal of the Royal Statistical
  Society: SERIES B: Statistical Methodology}, 821--849.

\bibitem[{Radicchi et~al.(2012)Radicchi, Fortunato, and
  Vespignani}]{radicchi2012citation}
Radicchi, F., Fortunato, S., and Vespignani, A. (2012), \enquote{Citation
  networks,} \textit{Models of science dynamics}, 233--257.

\bibitem[{Schecter and Quintane(2021)}]{schecter2021power}
Schecter, A. and Quintane, E. (2021), \enquote{The Power, Accuracy, and
  Precision of the Relational Event Model,} \textit{Organizational Research
  Methods}, 24, 802--829.

\bibitem[{Simmons and Hopkins(2005)}]{simmons2005constraining}
Simmons, B.~A. and Hopkins, D.~J. (2005), \enquote{The constraining power of
  international treaties: Theory and methods,} \textit{American Political
  Science Review}, 99, 623--631.

\bibitem[{Snijders and Lomi(2019)}]{snijders2019beyond}
Snijders, T.~A. and Lomi, A. (2019), \enquote{Beyond homophily: Incorporating
  actor variables in statistical network models,} \textit{Network Science}, 7,
  1--19.

\bibitem[{Therneau(2015)}]{therneau2015survival}
Therneau, T.~M. (2015), \textit{A Package for Survival Analysis in S}, version
  2.38.

\bibitem[{Therneau and Grambsch(2013)}]{therneau2013modeling}
Therneau, T.~M. and Grambsch, P.~M. (2013), \textit{Modeling survival data:
  extending the {C}ox model}, Springer Science \& Business Media.

\bibitem[{Verspagen(2007)}]{verspagen2007mapping}
Verspagen, B. (2007), \enquote{Mapping technological trajectories as patent
  citation networks: A study on the history of fuel cell research,}
  \textit{Advances in complex systems}, 10, 93--115.

\bibitem[{Vu et~al.(2017)Vu, Lomi, Mascia, and Pallotti}]{vu2017relational}
Vu, D., Lomi, A., Mascia, D., and Pallotti, F. (2017), \enquote{Relational
  event models for longitudinal network data with an application to
  interhospital patient transfers,} \textit{Statistics in medicine}, 36,
  2265--2287.

\bibitem[{Vu et~al.(2015)Vu, Pattison, and Robins}]{vpr-remslm-15}
Vu, D., Pattison, P., and Robins, G. (2015), \enquote{Relational event models
  for social learning in MOOCs,} \textit{Social Networks}, 43, 121--135.

\bibitem[{Zhou et~al.(2007)Zhou, Goldberg, Magdon-Ismail, and
  Wallace}]{zhou2007strategies}
Zhou, Y., Goldberg, M., Magdon-Ismail, M., and Wallace, A. (2007),
  \enquote{Strategies for cleaning organizational emails with an application to
  {E}nron email dataset,} in \textit{5th Conf. of North American Association
  for Computational Social and Organizational Science}, Pittsburgh: North
  American Association for Computational Social and Organizational Science.

\end{thebibliography}

\appendix

\section{Variation across samples}
\label{app:variation}

\begin{table}
\caption{\footnotesize \label{tab:resampling_rhem}Summary statistics (quantiles at probabilities $0.0,0.025,0.5,0.975,1.0$) of RHEM parameters over 100 samples.}
    \begin{center}
\begin{tabular}{lrrrrr}
  \hline
 & 0\% & 2.5\% & 50\% & 97.5\% & 100\% \\ 
  \hline
rec\_avg\_female & 0.183 & 0.196 & 0.212 & 0.230 & 0.236 \\ 
  rec\_avg\_senior & 0.322 & 0.327 & 0.339 & 0.354 & 0.358 \\ 
  rec\_avg\_legal & 0.105 & 0.111 & 0.131 & 0.153 & 0.165 \\ 
  rec\_avg\_trading & -0.124 & -0.122 & -0.108 & -0.092 & -0.089 \\ 
  send\_rec\_diff\_female & -0.203 & -0.200 & -0.185 & -0.172 & -0.166 \\ 
  send\_rec\_diff\_senior & -0.440 & -0.433 & -0.420 & -0.409 & -0.407 \\ 
  send\_rec\_diff\_dept & -0.753 & -0.751 & -0.737 & -0.724 & -0.717 \\ 
  rec\_set\_diff\_female & -0.286 & -0.259 & -0.203 & -0.133 & -0.091 \\ 
  rec\_set\_diff\_senior & -0.754 & -0.743 & -0.687 & -0.640 & -0.613 \\ 
  rec\_set\_diff\_dept & -1.072 & -1.056 & -0.987 & -0.929 & -0.903 \\ 
  exact\_repetition & -0.517 & -0.490 & -0.428 & -0.377 & -0.361 \\ 
  unoredered\_repetition & 0.962 & 0.970 & 1.019 & 1.076 & 1.087 \\ 
  rec\_sub\_rep\_1 & 0.020 & 0.022 & 0.028 & 0.036 & 0.038 \\ 
  rec\_sub\_rep\_2 & 0.321 & 0.374 & 0.468 & 0.564 & 0.580 \\ 
  rec\_sub\_rep\_3 & 1.258 & 1.358 & 1.868 & 2.347 & 2.588 \\ 
  rec\_sub\_rep\_4 & 3.047 & 3.270 & 4.943 & 7.006 & 8.248 \\ 
  send\_rec\_sub\_rep\_1 & 1.765 & 1.775 & 1.807 & 1.841 & 1.850 \\ 
  send\_rec\_sub\_rep\_2 & 4.078 & 4.502 & 5.173 & 6.083 & 6.269 \\ 
  send\_rec\_sub\_rep\_3 & 3.615 & 4.133 & 8.764 & 15.288 & 16.344 \\ 
  reciprocation & 0.555 & 0.576 & 0.607 & 0.640 & 0.652 \\ 
  out\_in\_pop & -0.002 & 0.001 & 0.010 & 0.018 & 0.019 \\ 
  interact\_rec\_1 & 2.365 & 2.401 & 2.599 & 2.798 & 2.866 \\ 
  interact\_rec\_2 & 4.325 & 5.322 & 7.851 & 10.518 & 10.962 \\ 
  interact\_rec\_3 & 3.922 & 10.185 & 22.745 & 33.437 & 36.689 \\ 
  in\_balance & 0.051 & 0.057 & 0.068 & 0.076 & 0.078 \\ 
  out\_balance & -0.125 & -0.123 & -0.111 & -0.098 & -0.093 \\ 
  transitive\_closure & 0.045 & 0.048 & 0.057 & 0.073 & 0.073 \\ 
  cyclic\_closure & -0.061 & -0.057 & -0.042 & -0.025 & -0.022 \\ 
   \hline
\end{tabular}
    \end{center}
\end{table}

We repeated parameter estimation on 100 different samples of non-events associated with each observed event. In Table~\ref{tab:resampling_rhem} we report summary statistics of RHEM parameters over these 100 samples and in Table~\ref{tab:resampling_dyadic} we report summaries of parameters of the dyadic REM. We can observe that the distributions of parameters capturing (sender-specific) subset repetition and interaction among receivers have a dispersion that is increasing with the order of these covariates. This can be explained by the fact that these covariates are increasingly sparse among the non-events, that is, most randomly sampled non-events assume the value zero on these higher-order statistics. As indicated in the conclusion, this sparsity implies the need for a large sample size when parameters are estimated via case-control sampling -- but could possibly be reduced by other parameter estimation techniques (for instance, via stratified sampling or MCMC methods). However, even the distributions of these parameters consistently do not cross zero. Parameter distributions of most other effects show less relative variability.

\begin{table}
\caption{\footnotesize \label{tab:resampling_dyadic}Summary statistics (quantiles at probabilities $0.0,0.025,0.5,0.975,1.0$) of dyadic REM parameters over 100 samples.}
    \begin{center}
\begin{tabular}{lrrrrr}
  \hline
 & 0\% & 2.5\% & 50\% & 97.5\% & 100\% \\ 
  \hline
rec\_avg\_female & 0.224 & 0.237 & 0.254 & 0.275 & 0.280 \\ 
  rec\_avg\_senior & 0.448 & 0.457 & 0.468 & 0.485 & 0.489 \\ 
  rec\_avg\_legal & 0.119 & 0.129 & 0.151 & 0.174 & 0.182 \\ 
  rec\_avg\_trading & -0.240 & -0.237 & -0.221 & -0.202 & -0.197 \\ 
  send\_rec\_diff\_female & -0.252 & -0.251 & -0.234 & -0.220 & -0.215 \\ 
  send\_rec\_diff\_senior & -0.512 & -0.504 & -0.491 & -0.479 & -0.476 \\ 
  send\_rec\_diff\_dept & -0.909 & -0.905 & -0.892 & -0.878 & -0.867 \\ 
  rec\_sub\_rep\_1 & 0.036 & 0.037 & 0.044 & 0.052 & 0.054 \\ 
  send\_rec\_sub\_rep\_1 & 2.505 & 2.508 & 2.530 & 2.547 & 2.555 \\ 
  reciprocation & 0.976 & 0.985 & 1.007 & 1.034 & 1.042 \\ 
  out\_in\_pop & 0.008 & 0.011 & 0.018 & 0.026 & 0.028 \\ 
  in\_balance & 0.081 & 0.088 & 0.098 & 0.105 & 0.110 \\ 
  out\_balance & -0.059 & -0.059 & -0.048 & -0.037 & -0.032 \\ 
  transitive\_closure & 0.102 & 0.105 & 0.116 & 0.127 & 0.134 \\ 
  cyclic\_closure & -0.126 & -0.122 & -0.107 & -0.094 & -0.086 \\ 
   \hline
\end{tabular}
    \end{center}
\end{table}

\end{document}